\documentclass[rmp,aps,twocolumn,floatfix,groupedaddress,superscriptaddress,amsmath,a4paper,twoside,showkeys]{revtex4}
\usepackage{graphicx,txfonts}
\usepackage{chemarr}
\usepackage{times,longtable}
\usepackage{hyperref}
\usepackage{version}
\usepackage{color}
\graphicspath{{figures/},{figures/pdf/},{figures/eps/}}

\bibpunct{[}{]}{,}{n}{}{,}

\providecommand{\addhyphen}[1]{#1.---}

\makeatletter
\renewcommand \paragraph{%
  \@startsection
    {paragraph}%
    {4}%
    {\parindent}%
    {\z@}%
    {-.1em}%
    {\normalfont\normalsize\itshape\addhyphen}%
}%
\makeatother

\begin{document}

\title{
Computing fluxes and chemical potential distributions in
  biochemical networks: energy balance analysis of the human red blood
  cell}

\author{Daniele De Martino}

\affiliation{Dipartimento di Fisica, Sapienza Universit\`a di Roma, P.le A. Moro 2, Roma, Italy }

\author{Matteo Figliuzzi}

\affiliation{Dipartimento di Fisica, Sapienza Universit\`a di Roma, P.le A. Moro 2, Roma, Italy }

\author{Andrea De Martino}

\affiliation{CNR/IPCF, Dipartimento di Fisica, Sapienza Universit\`a di Roma, P.le A. Moro 2, Roma, Italy }

\author{Enzo Marinari}

\affiliation{Dipartimento di Fisica, Sapienza Universit\`a di Roma, P.le A. Moro 2, Roma, Italy }


\begin{abstract}
The analysis of non-equilibrium steady states of biochemical
  reaction networks relies on finding the configurations of fluxes and
  chemical potentials satisfying stoichiometric ({\it mass balance})
  and thermodynamic ({\it energy balance}) constraints. Efficient
  methods to explore such states are crucial to predict reaction
  directionality, calculate physiologic ranges of variability,
  estimate correlations, and reconstruct the overall energy balance of
  the network from the underlying molecular processes. While different
  techniques for sampling the space generated by mass balance
  constraints are currently available, thermodynamics is generically
  harder to incorporate. Here we introduce a method to sample the free
  energy landscape of a reaction network at steady state. In its most
  general form, it allows to calculate distributions of fluxes and
  concentrations starting from trial functions that may contain prior
  biochemical information. We apply our method to the human red blood
  cell's metabolic network, whose space of mass-balanced flux states
  has been sampled extensively in recent years. Specifically, we
  profile its thermodynamically feasible flux configurations,
  characterizing in detail how fluctuations of fluxes and potentials
  are correlated. Based on this, we derive the cell's energy balance
  in terms of entropy production, chemical work done and thermodynamic
  efficiency.
\end{abstract}


\maketitle

\section{Introduction}

The dynamics of a chemical reaction network is ruled by the underlying
thermal fluctuations through the Arrhenius law that relates the rates
of reactions to the activation energies and the strength of noise (the
temperature). When non-equilibrium steady states (NESS) are reached,
the net flow of reactions is constrained to proceed downhill in the
free energy landscape \cite{dedo,bq1}. Many aspects involved in the
analysis of biochemical networks (like the cellular metabolism of
living organisms) at stationarity hinge on the explicit inclusion of
thermodynamic constraints into the network model; among them, the
assignment of reaction directions (and in turn the calculation of
feasible flux configurations), the assessment of metabolite
producibility, the prediction of metabolite concentrations and the
estimation of chemical potentials.

Steady-state schemes used to predict flux values are mostly based on
mass-balance constraints only, and the way in which thermodynamics is
included could significantly impact their results. For
Flux-Balance-Analysis (FBA) \cite{kau,palrev}, where mass-balanced
flux configurations are collapsed into a single optimal solution that
maximizes a pre-determined objective function \cite{sauer,biomass},
the removal of thermodynamic inconsistencies by additional energetic
constraints has been proven to be useful to estimate concentrations
and reaction affinities besides fluxes
\cite{beard1,beard2,kummel2,hoppe,tbmfa}. In absence of clear
optimization criteria or, more generally, to retrieve global
information on the feasible network states allowed in given
extracellular and intracellular conditions (e.g. physiologic ranges of
variability), it is instead important to characterize the space of
flux configurations compatible with mass- and energy-balance
constraints in a statistically robust way, as allowed e.g. by Monte
Carlo (for small networks \cite{palsson}) or message-passing (for
larger networks \cite{zecch}) methods for mass-balance equations.

Here we propose a scalable technique to obtain refined information on
the {\it distribution} of fluxes, chemical potentials and
intracellular concentrations for non-equilibrium steady states, as
well as predictions for correlations and reaction directions. The
method is based on perceptron learning, extends the sampling
procedure used in \cite{pnas} to explore the space of flux states
compatible with minimal stability constraints {\it \`a la Von Neumann}
\cite{jstat} and uses the stoichiometric matrix and the experimentally
determined potentials in standard conditions of a set of metabolites
as its basic input. In essence, it generates feasible configurations
of fluxes and concentrations by exploiting the input information and a
suitably chosen update dynamics to build-up {\it correlations} between
chemical potentials and fluxes. We apply it to the control case of the
model of the human red cell (hRBC) metabolism reconstructed in
\cite{palsson}, for which we derive and analyze the viable state space
and the emerging correlations. This ultimately allows to characterize
the cell as a chemical engine. In particular, we shall appraise its
chemical energy balance in the standard terms of entropy production,
work done and thermodynamic efficiency.

\section{Background}

A metabolic reaction network is defined by a matrix $\boldsymbol{\Xi}$
encoding the stoichiometric coefficients $\xi_i^\mu$ of the compounds
$\mu=1,\ldots,M$ in the reaction $i=1,\ldots,N$. Conventionally,
positive (resp. negative) stoichiometric indices characterize the
products (resp. substrates) of the forward reaction.  Intakes
(resp. outtakes) are characterized as having only positive
(resp. negative) stoichiometry. Neglecting molecular noise and
dilution due to expanding cell volume, the concentration $c^\mu$ of
metabolite $\mu$ evolves according to
\begin{equation}
\dot{c}^\mu  = \sum_{i=1}^N \xi_i^\mu \nu_i(\mathbf{c},\mathbf{k},\ldots)\;,
\end{equation}
where $\nu_i$ is the net flux of reaction $i$, which depends in
principle on the concentration vector $\mathbf{c}$, on a vector of
reaction constants $\mathbf{k}$, as well as on other factors like
enzyme availability and kinetics, transport details, etc. 
The usual working hypothesis is to assume stationarity: in this case
the flux vector $\boldsymbol{\nu}$
satisfies the $M$ linear mass-balance equations (MBE)
\begin{equation}\label{mb}
\boldsymbol{\Xi\nu}=\mathbf{0}\;.
\end{equation} 
Bounds of the type $\nu_i^{\text{min}}\leq\nu_i\leq\nu_i^{\text{max}}$
are also usually specified. These include both physical
considerations, like an assignment of reversibility by thermodynamic
arguments, and functional aspects, e.g. the fact that a certain
reaction should operate above a given flux threshold in order to
provide a physiological function. Cellular metabolic networks usually
have $N>M$ so that the system (\ref{mb}) is underdetermined, leading
to a solution space of dimension $D=N-\text{rank}
(\boldsymbol{\Xi})$. Uniform sampling is computationally affordable
only for small enough $D$ (a few tens), e.g. via Monte Carlo methods
\cite{palsson}. However if the interest is focused on finding flux
configurations that are optimal with respect to specific biological
functionalities (e.g. biomass production), the solutions of (\ref{mb})
can be further constrained to maximize an objective function. This is
the standard framework of FBA, as implemented with considerable
success to describe for example optimal bacterial growth in wild type
and knock-out conditions \cite{eip,moma,room}.

In a different approach one leaves the solution space
functionally unconstrained while studying the metabolite production
profiles compatible with a given extracellular medium. For any choice
of the environment (i.e. of the intakes, the outtakes being left
unspecified), a metabolite $\mu$ is producible if there exists a flux
vector $\boldsymbol{\nu}$ such that $\sum_i\xi_i^\mu\nu_i>0$
\cite{belta}. Feasible production profiles are then described by the
solutions of
\begin{equation}\label{vn}
\boldsymbol{\Xi\nu}\geq\mathbf{0}\;.
\end{equation} 
The resulting vector $\mathbf{y}\equiv\boldsymbol{\Xi\nu}$ encodes the
information on whether each metabolite is producible ($y^\mu>0$) or
not ($y^\mu=0$) in a given feasible flux state. Producible metabolites
might either be employed in other macromolecular processes or become
cellular outtakes, so that ``objective functions'' can in this way
emerge as statistically robust production profiles
\cite{kyoto}. Eq. \ref{vn} is Von Neumann's flux stability constraint
(VNC) for production networks \cite{jvn} and simply states that for
each chemical species at stationarity the overall consumption cannot
exceed the total supply. The inequality that distinguishes (\ref{vn})
from (\ref{mb}) allows for the efficient sampling of its solution
space even for genome-scale networks of the size of E. coli, where $D$
is of the order of a few hundreds \cite{pnas}.

Going beyond the flux problem, the second law of thermodynamics
dictates that in a NESS the reaction directions
$s_i\equiv\text{sign}(\nu_i)$ must be related to the chemical driving
forces (affinities) $\Delta G_i$ by
\begin{equation}\label{tc}
s_i \Delta G_i \leq 0~~~~~\forall i\;,
\end{equation} 
where the equality only holds if $i$ is in equilibrium. The free energy changes
$\Delta G_i$ can be written in terms of the chemical potentials
$g^\mu$ (the Gibbs free energy per mole of the species $\mu$) as
\begin{equation}
\Delta G_i = \sum_{\mu=1}^M \xi_i^\mu g^\mu \;.
\end{equation}
The existence of non-trivial states of chemical equilibrium (trivial
states are the ones with $g^\mu=0$ for each $\mu$) implies $M >
\text{rank}(\boldsymbol{\Xi})$. The thermodynamic constraints
(\ref{tc}) are usually implemented a priori, by pre-assigning reaction
reversibilities based on the estimation of chemical potentials in
physiologic conditions \cite{kummel, fleming}. In other words, the
flux of a reaction classified as irreversible in the forward direction
should be sampled under the condition of being non-negative. In cases
of mis-assignments, the flux problem can lead to inconsistent
configurations \cite{bq1}. Note that, in absence of prescribed flux
bounds, given a vector $\mathbf{z}=\{z_i\}$ ($i=1,\ldots,N$) such that
\begin{equation}
\sum_{i\in R\setminus U} \xi_i^\mu z_i=0 ~~~~~\forall\mu\;, 
\end{equation}
where the sum extends over all reactions (set $R$) excluding uptakes
(set $U$), for each solution $\boldsymbol{\nu}^\star=\{\nu_i^\star\}$
of the flux problem (\ref{mb}) or (\ref{vn}), the vector defined as
$\boldsymbol{\nu}^\star+k \mathbf{z}$ is again a solution for any
$k\in\mathbb{R}$. In essence, a thermodynamically consistent assignment of reaction
directions reduces this degeneracy by bounding some degrees of
freedom. A concrete example of this is discussed in Appendix A1. Our goal in this note is to devise a tool to obtain flux vectors $\boldsymbol{\nu}=\{\nu_i\}$ and chemical
potential vectors $\mathbf{g}=\{g^\mu\}$ that are joint solutions of
(\ref{vn}) (or (\ref{mb})) and (\ref{tc}).

\section{Methods}

We have considered two (non-equivalent) solution schemes. In Method
(a), the flux and the chemical potential problems are decoupled:
first, the former (i.e. Eq. (\ref{mb}) or (\ref{vn})) is solved for
$\boldsymbol{\nu}$ for a given reversibility assignment (e.g. based on
physiological data), then (\ref{tc}) is solved for $\mathbf{g}$ using
as an input the signs $s_i$ of the net fluxes that solve the flux
problem. This procedure allows to remove thermodynamically
inconsistent solutions of the flux problem and provides estimates for
chemical potentials, but it doesn't allow to predict reaction
directionalities. Method (b) instead solves the mass balance and the
energy balance problems jointly without any prior assumption on
reversibility. We shall begin this section by outlining the two
algorithms; then we shall briefly discuss the algorithmic setup and
the cellular data we have analyzed (a detailed account of these issues
is presented in Appendices A2 and A3.

\subsection{Method (a) Decoupling the flux and the free energy problems}

The solutions of the flux problems (\ref{mb}) and (\ref{vn}) (starting
from a priori reversibility assignments) can be sampled respectively
by Monte Carlo methods (for small networks, see e.g. \cite{palsson}) and
by the MinOver$^+$ scheme (even for large systems, see
e.g. \cite{pnas,andrea}). We assume to have obtained the resulting
distributions for the net fluxes and hence a vector
$\mathbf{s}=\{s_i\}$ of net reaction directions ($+1$ for forward,
$-1$ for backward, $0$ for bidirectional). The free energy landscape
reconstruction problem consists, given $\mathbf{s}$, in sampling
vectors $\mathbf{g}$ that satisfy (\ref{tc}), namely such that
\begin{equation}\label{constr}
x_i\equiv -s_i \sum_{\mu=1}^M \xi_i^\mu g^\mu \geq 0~~~~~\forall i\;.
\end{equation}
To approach it, we note that the problems (\ref{constr}) and
(\ref{vn}) are formally equivalent. By analogy with \cite{jstat}, we
can then employ the MinOver scheme originally designed to analyze
perceptron learning \cite{mez} and later adapted to generate solutions
of (\ref{vn}). Let $P_0(\mathbf{g})\equiv \prod_{\mu=1}^M
P_0^\mu(g^\mu)$ denote a trial probability distribution of chemical
potentials that contains some a priori information on
experimentally determined potentials and concentrations. For instance,
$P_0^\mu$ could be a uniform distribution centered around the known
experimental values of $g^\mu$ and of sufficiently large width to span
several orders of magnitude in trial concentrations, or a flat
unbiased uniform distribution. In essence, the algorithm is designed
to modify $P_0$ by building up correlations between $g^\mu$'s, until a
distribution matching the solution space of (\ref{constr}) is
achieved. The steps are as follows.
\begin{enumerate}
\item Generate a chemical potential vector $\mathbf{g}=\{g^\mu\}$ from
  $P_0(\mathbf{g})$.
\item Compute $\mathbf{x}=\{x_i\}$ from (\ref{constr}) and
  $i_0=\text{arg }\min_i x_i$.
\item If $x_{i_0}\geq 0$ then $\mathbf{g}$ is a thermodynamically
  consistent chemical potential vector for $\mathbf{s}$; exit (or go
  to 1 to obtain a different solution).
\item If $x_{i_0}<0$, update $\mathbf{g}$ as 
\begin{equation}\label{deltag}
g^\mu~\to~g^\mu-\alpha s_{i_0} \xi_{i_0}^{\mu}
\end{equation}
(where $\alpha>0$ is a constant), go to 2 and iterate until convergence.
\end{enumerate}
As is generally true in MinOver schemes, the reinforcement term in
(\ref{deltag}) drives the gradual adjustment of potentials by ensuring
that, at every iteration, the least satisfied constraint (labeled
$i_0$) is improved. In particular, the chemical potentials of
metabolites that are produced (resp. consumed) in reaction $i_0$ are
decreased (resp. increased) at each time step until a state is
achieved where all unidirectional fluxes descend in the free energy
landscape. Note that even though the constraint (\ref{constr}) is
absent if $s_i=0$, the above method still allows to retrieve
information about the chemical potential of a metabolite involved in
reversible processes (unless it is {\it only} involved in reversible
processes, in which case its $g^\mu$ is never updated). Convergence to
a solution (if any) is guaranteed for any $\alpha>0$, the proof being
a minor modification of the one shown in \cite{jstat}, and the space
of feasible chemical potentials can be sampled by re-starting the
process from different chemical potential vectors belonging to
$P_0(\mathbf{g})$. In this way, the final outcome is a set of
correlated probability distributions for the $g^\mu$'s. Note that, at
odds with the method proposed in \cite{kummel2}, the final chemical
potentials can exceed the bounds defined by $P_0(\mathbf{g})$. Details
about the implementation of this scheme (e.g. the choice of $\alpha$)
are described in Appendix A3.

\subsection{Method (b) Joint solution of the flux and free energy problem}

Method (b) aims at solving the flux and the free energy problems
simultaneously without relying on a priori information on reaction
reversibility (i.e. all are initially assumed to be
bidirectional). Besides a trial distribution $P_0(\mathbf{g})$ for the
chemical potentials, it uses the chemical potentials $g^\mu_{(ext)}$
of the metabolites subject to uptakes. We focus on the case where the
flux problem is represented by the VNC (\ref{vn}). From a
computational viewpoint, sampling the solution space of (\ref{vn}) is
done more conveniently by introducing a parameter $\rho>0$ such that
the system (\ref{vn}) is recovered for $\rho\to 1$ (see
e.g. \cite{jstat,pnas}). In a fully reversible setting, the VNC takes a
simple form by writing $\boldsymbol{\Xi}=\mathbf{A}-\mathbf{B}$ where
$\mathbf{A}=\{a_i^\mu\}$ and $\mathbf{B}=\{b_i^\mu\}$ denote,
respectively, the matrices of output and input stoichiometric
coefficients, and by re-defining the flux variable $\nu_i$ as
\begin{equation}\label{nu}
\nu_i\equiv s_i\phi_i\;,\;\;\mbox{where}\;\;s_i\equiv
\text{sign}(\nu_i)\;\mbox{and}\;\phi_i\equiv |\nu_i|\;.
\end{equation}
Introducing the shorthand
\begin{equation}
\xi_i^\mu(\rho,\nu_i)=
\theta(\nu_i)(a_i^\mu-\rho b_i^\mu)+\theta(-\nu_i)(b_i^\mu-\rho a_i^\mu)
\end{equation}
one sees \cite{matteofi} that the solutions of (\ref{vn}) correspond to those of
\begin{equation}\label{yc}
y^\mu(\rho)=\sum_{i=1}^N\xi_i^\mu(\rho,\nu_i)\phi_i\geq 0~~~~~\forall\mu\;,
\end{equation}
for $\rho\to 1$. The algorithm proceeds as follows.
\begin{enumerate}
\item Initialize $\rho=\overline{\rho}<1$ (e.g. $\overline{\rho}=0$).
\item Generate a chemical potential vector $\mathbf{g}=\{g^\mu\}$ from
  $P_0(\mathbf{g})$.
\item Assign reaction directions
\begin{enumerate}
\item[3A.] For intracellular reactions: compute affinities as $\Delta
  G_i=\sum_\mu\xi_i^\mu g^\mu$, and assign directions as
  $s_i=-\text{sign}(\Delta G_i)$
\item[3B.] For intakes: if $g^\mu<g^\mu_{(ext)}$ then $s_i=1$ (intake
  is active), otherwise $s_i=0$ (inactive).
\end{enumerate}
\item Generate a vector $\boldsymbol{\phi}=\{\phi_i\}$ from a uniform distribution, e.g. in $[0,1]$. ($\phi_i$ is the absolute values of flux $\nu_i$, see (\ref{nu}).)
\item Compute $\mathbf{y}(\overline{\rho})=\{y^\mu(\overline{\rho})\}$ from (\ref{yc}) and $\mu_{0} =\text{arg } \min_\mu y^\mu(\overline{\rho})$.
\item If $y^{\mu_0}(\overline{\rho})\geq 0$, then the configuration $(\boldsymbol{\nu},\mathbf{g})$ with $\nu_i=s_i\phi_i$ is a thermodynamically feasible configuration of fluxes and chemical potentials for $\rho=\overline{\rho}$ ; go to 8.
\item If $y^{\mu_0}(\overline{\rho})< 0$, then update $\boldsymbol{\phi}$ as 
\begin{equation}
\phi_i~\to~\phi_i'=\phi_i+\beta \xi_i^{\mu_0}
\end{equation}
with $\beta>0$ a constant.
If $\text{sign}(\phi_i')<0$, then update $\mathbf{g}$ as 
\begin{equation}
g^\mu~\to~ g^\mu+\alpha s_i \xi_i^\mu
\end{equation}
and re-assign directions as in step 3. Let $\{s_i'\}$ denote the updated directions. Then
\begin{itemize}
\item If $s_i'=s_i$, then set $\phi_i=0$ and go to 5 (iterate until convergence)
\item If $s_i'=-s_i$, then set $\phi_i=|\phi_i'|$ and go to 5 (iterate until convergence to a solution)
\end{itemize}
\item If $\overline{\rho}\leq\rho_{max}=1-\epsilon$, update $\overline{\rho}$ as $\overline{\rho}\to\overline{\rho}+\delta$ with $\delta>0$ and go to $5$ and iterate until convergence
\end{enumerate}
The idea behind this scheme is to mimic a relaxational dynamics into a
thermodynamically stable steady state. Following initialization, in
step 3 directions are assigned according to a chemical potential
vector. This assignment satisfies (\ref{tc}) by definition and
encodes information on the connected correlations
among reaction affinities, i.e.
\begin{equation}
\langle \Delta G_i \Delta G_j \rangle_c =  \sum_{\mu=1}^M \xi_i^\mu  \xi_j^\mu (g^\mu-\langle g^\mu\rangle)^2\;,
\end{equation}    
$\langle g^\mu\rangle$ denoting the average of $g^\mu$ according to
the trial function $P_0$. The algorithm then performs a MinOver scheme
for fluxes, trying to find a feasible flux configuration compatible
with this assignment. As the dynamics proceeds, some of the fluxes
might display a preference to revert. In these cases, we let the
chemical potential vector slowly evolve following the scheme of Method
(a). If this provides a new set of directions agreeing with the
inversion, the latter is accepted, otherwise the reaction is shut down
and the process is iterated. Once a configuration of fluxes and
chemical potentials that satisfy (\ref{tc}) and (\ref{yc}) at
$\rho=\overline{\rho}$ is achieved, $\rho$ is increased until
$\rho=1-\epsilon$ (with $\epsilon$ the desired precision, typically
$10^{-3}$ or less). Results at $\rho=1$ can then easily be
extrapolated, and different solutions at $\rho=1$ can be sampled by
re-starting from step 1 with a different vector $\mathbf{g}$ from
$P_0$. In this way, flux configurations evolve in a thermodynamically
coherent fashion till they reach production feasibility. During
evolution, the algorithm modifies the initial distribution of chemical
potentials exploring states close to the initial conditions but with a
dynamics that correlates them according to feasible flux
directions. Details about the implementation of Method (b) are
reported in Appendix A3.

\subsection{The hRBC metabolic network}

We shall apply our schemes to the cellular metabolic network of the
hRBC, reconstructed in \cite{palsson}. The detailed reconstruction as
well as the auxiliary data employed for this study are exposed in
Section S2. It consists of $35$ intracellular reactions among $39$
metabolites subject to $12$ uptakes.  The network includes three
pathways, namely glycolysis (reactions 1--13) and the pentose
phosphate pathway (reactions 14--21), through which glucose is
degraded to produce high-energy molecules needed to maintain osmotic
pressure, transmembrane potential and redox state, plus a nucleotide
salvage pathway (reactions 22--32). The functional part of the network
lies in the three reactions (33--35) that are performing chemical
work, namely: the ATPase pump, that maintains osmotic pressure and
transmembrane potential by exchanging Na$^+$ and K$^+$ ions with the
surrounding plasma; the NADHase pump, carrying out the reduction of
oxidized heme groups; and the NADPHase pump, whose function is to
reduce the glutathione (GSH) molecules that continuously get oxidized
while acting against free radicals.

The main reason for focusing on this network lies in the fact that for this case it is possible to carry out a full comparison with previous results obtained by sampling the solution space of both (\ref{mb}) \cite{palsson} and (\ref{vn}) \cite{andrea}, as well as with experimental estimates of concentration profiles and chemical potentials. The information on net reaction directions required for Method (a) has been extracted from these studies (see Section S3). As for the trial distribution of chemical potentials $P_0(\mathbf{g})=\prod_\mu P_0^\mu(g^\mu)$ required by both Methods (a) and (b), we have considered two different cases (explained in detail in Section S3). 

(i) {\it Poor input information}. By virtue of (\ref{tc}) a certain amount of a priori information on the $g^\mu$'s (to be incorporated in $P_0$) is required in order to reconstruct the entire free energy landscape. Case (i) reproduces a situation in which this prior knowledge roughly matches the necessary minimum. Specifically, the chemical potentials of $8$ metabolites, namely the four sources that are present in the hRBC network, i.e. glucose (GLC), lactate (LAC), NH$_3$ and CO$_2$, plus inosine monophosphate (IMP), adenine (ADE), NAD and NADP, is fixed in $P_0$ and the corresponding variables are not updated. For the other compounds we have chosen a $P_0^\mu$ that reproduces the overall statistics of chemical potentials: specifically, each $g^\mu$ (in units of KJ/mol) is selected independently and uniformly in $[0,2000]$ with probability $p=0.2$ and in $[-5000,0]$ with probability $1-p=0.8$. 

(ii) {\it Rich input information}. In this case $P_0(\mathbf{g})$ includes knowledge on the chemical potential of metabolites whose intracellular concentrations $c^\mu$ are experimentally known (with errors), extracted from the formula $g^\mu=g_0^\mu+RT\log c^\mu$. However, contrary to case (i) no chemical potential is fixed with the exception of that H$_2$O, assuming water to be in a condensed phase. When we were unable to find reliable concentration estimates, the trial distribution of $\log c^\mu$ was taken to be uniform, centered in $-4$ and spanning four values symmetrically around the mean.

\section{Results}  
 
\subsection{Method (a), poor input information}

\begin{figure}
\begin{center}
\includegraphics*[width=.45\textwidth]{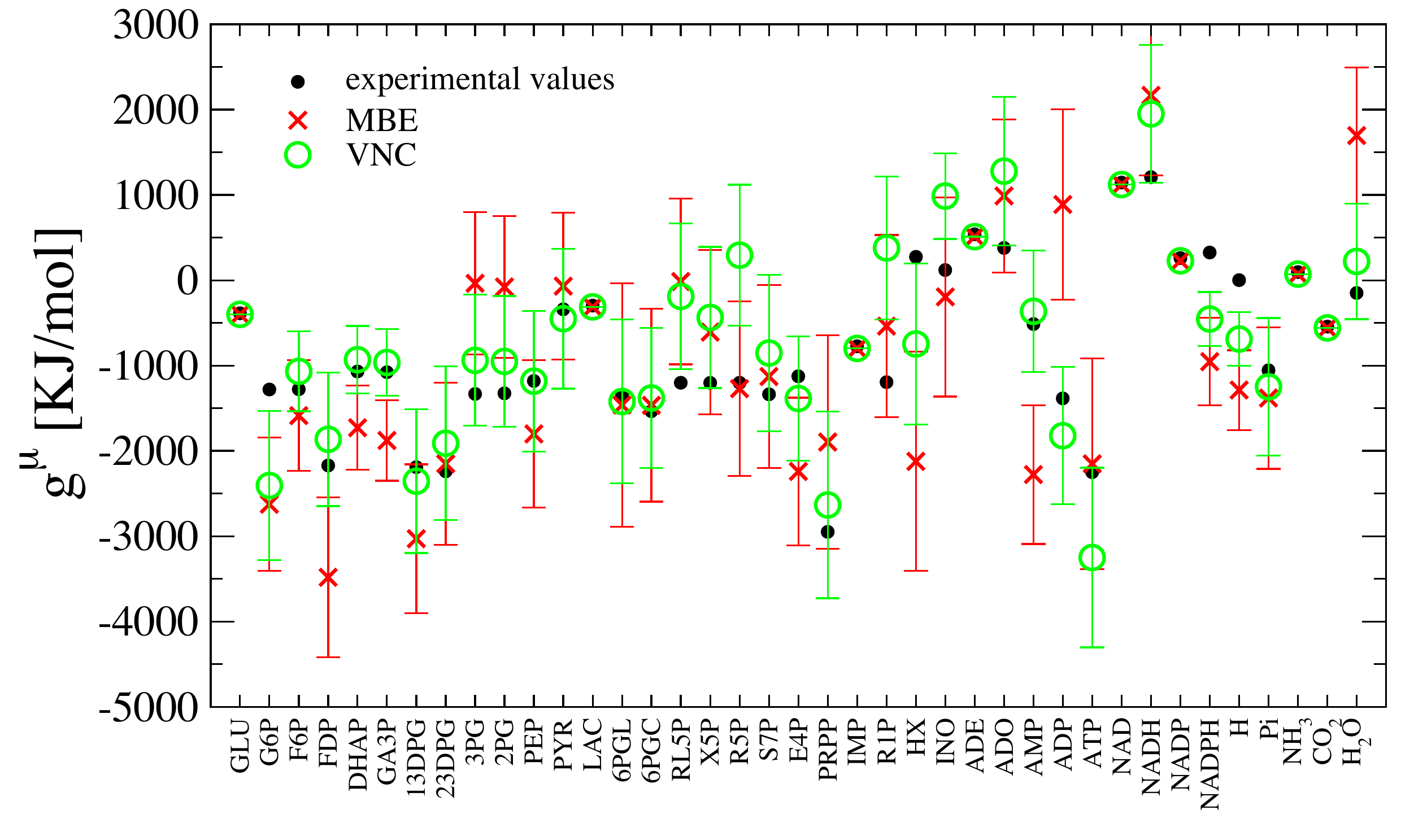}
\caption{Chemical potentials of metabolites in the hRBC metabolic network computed by Method (a) with poor input information: experimental estimates (black markers) and values obtained for MBE (red markers) and VNC (green markers) direction assignments. The initial spread of chemical potentials encoded in $P_0$ (spanning either the $[0,2000]$ or the $[-5000,0]$ range) is not shown. The errors on the experimental estimates of the fixed chemical potentials fall within the size of the dots.}
\label{fig1}
\end{center}
\end{figure}

Results for the chemical potential landscape of the hRBC reconstructed by Method (a) from poor input information and from the reaction directions obtained in \cite{palsson} and \cite{andrea} for MBE and VNC respectively are shown in Fig.~\ref{fig1}. 
Both mass-balance conditions turn out to provide thermodynamically feasible configurations of reaction directions and in both cases one obtains a broad agreement between computed and experimentally measured chemical potentials. This shows that Method (a), apart from being able to verify the thermodynamic feasibility of a flux state, is able to retrieve information on the energy landscape. Keeping in mind that MBE represent tighter constraints on fluxes than VNC and that the networks used in \cite{palsson} and \cite{andrea} are not identical (specifically in uptakes), one sees that the directions extracted from (\ref{vn}) slightly outperform those resulting from (\ref{mb}) in predicting chemical potentials, while both fail to predict (via Method (a)) the $g^\mu$'s of key metabolites like NADH and NADPH. The magnitude of the error bars reflects the considerable initial uncertainty we have assumed in the input information, as prior knowledge in this case is the minimal needed to solve (\ref{tc}) unambiguously. 
 
\subsection{Method (a), rich input information}

\begin{figure}
\begin{center}
\includegraphics*[width=0.45\textwidth]{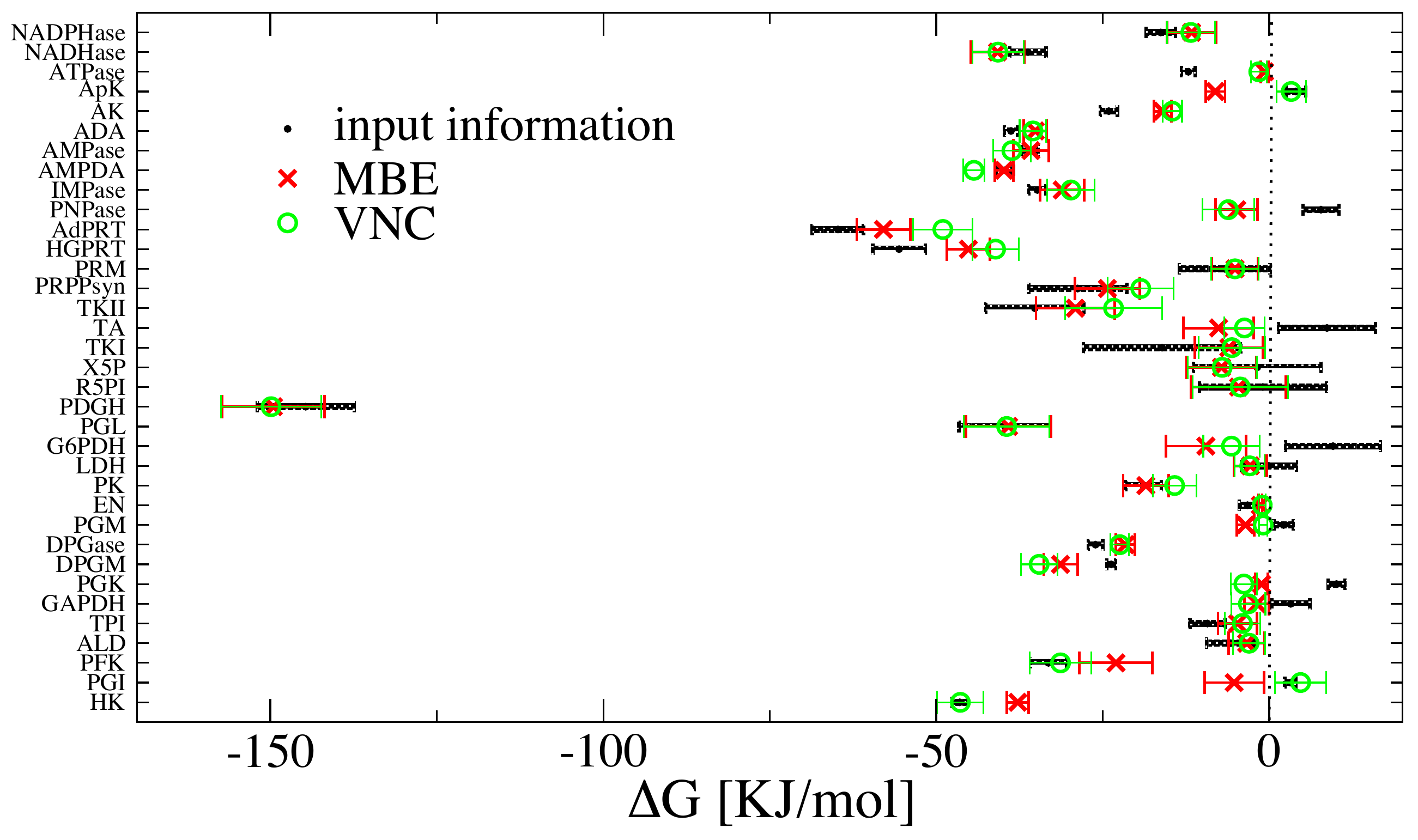}
\includegraphics*[width=0.45\textwidth]{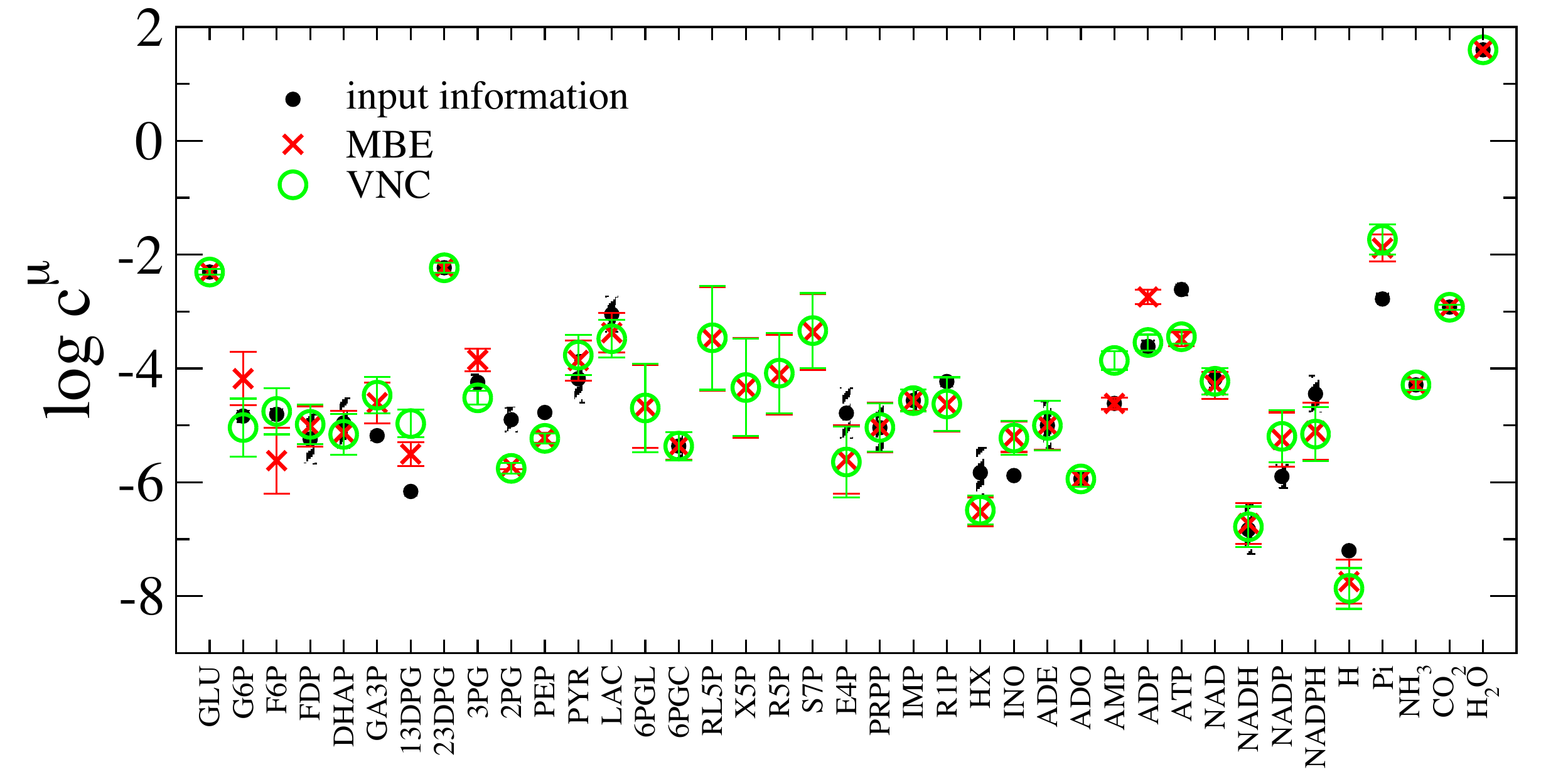}
\caption{Predictions obtained for the hRBC network by Method (a) with rich input information. Top: free energy changes: input information with error (black markers) and values obtained for MBE (red markers) and VNC (green markers) direction assignments. Bottom: measured metabolite (log-)concentrations: input information with error (black markers) and values obtained for MBE (red markers) and VNC (green markers) direction assignments.}
\label{fig2}
\end{center}
\end{figure}

A much refined prediction is obtained using a rich input information. Results are displayed in Fig.~\ref{fig2}. 
The computed affinities display significant changes with respect to the picture embedded in $P_0$. For instance, several of the initial bounds on affinities indicate that the free energy change in a reaction is positive (e.g. PGI, PGK, G6PDH, TA), due either to the actual experimental estimates for the affinities or to the initial uncertainty we place on concentrations. Such bounds are altered by MinOver in a direction compatible with physiologic assignments (see Table S2), suggesting that the implementation of Method (a) correlates the fluctuations of chemical potentials. At the same time, we observe that in some cases fluctuations of $g^\mu$ exceed the initial boxes defined by $P_0$. Since the resulting ranges are compatible with the intake configuration, they may provide an estimate of the statistical fluctuations from cell to cell. This also allows to obtain an estimate for the concentration range, including for metabolites whose level has not been experimentally probed (6PGL, RL5P, X5P, R5P, S7P). Our predictions for the levels of (1,3)-diphosphoglycerate ($(1,3)$-DPG), 2-phosphoglycerate (2PG) and phosphoenolpyruvate (PEP) differ from the experimental estimates. This is a consequence of the fact that we are forcing the phosphoglycerate kinase (PGK) and the glyceraldehyde phosphate dehydrogenase (GAPDH) reactions in the forward direction, in agreement with the steady state direction assignments for glycolysis, even if the experimental values would classify them as reversible. In addition, we obtain different levels of key metabolites like ATP and inorganic phosphate, while our predictions for ADP and AMP fail in the MBE and VNC conditions, respectively. This can be partially traced back to the difficulties inherent in assessing the potentials of highly exchanged metabolites \cite{fleming}. On the other hand, the experimental estimates of these concentrations vary considerably across the literature (e.g. \cite{beutler} \emph{vs} \cite{leninger}). Finally, a net intake of phosphate groups and a net outtake of CO$_2$ is predicted both for the MBE and the VNC states even if in physiological conditions the ratio of internal concentrations and the external levels in the blood would suggest  otherwise (see Tables S1 and S3). This, together with a slight inconsistency in the pH level, seems to call for the inclusion in the network of the carbonic anhydrase (CA-I) reaction, as well as of a bicarbonate (HCO$_3^-$) outtake. The former carries out the intracellular pH buffering by hydrating CO$_2$ into bicarbonate, while the latter releases HCO$_3^-$ through the BAND3 membrane protein, which can cover up to 25\% of the membrane surface \cite{band3}. So, even if in the standard biochemical state species that differ only by the hydration level are usually lumped together in the reactants list, we have chosen to treat carbon dioxide and bicarbonate as different metabolites in the implementation of Method (b).

\subsection{Method (b), rich input information}

\begin{figure}
\begin{center}
\includegraphics*[width=0.45\textwidth]{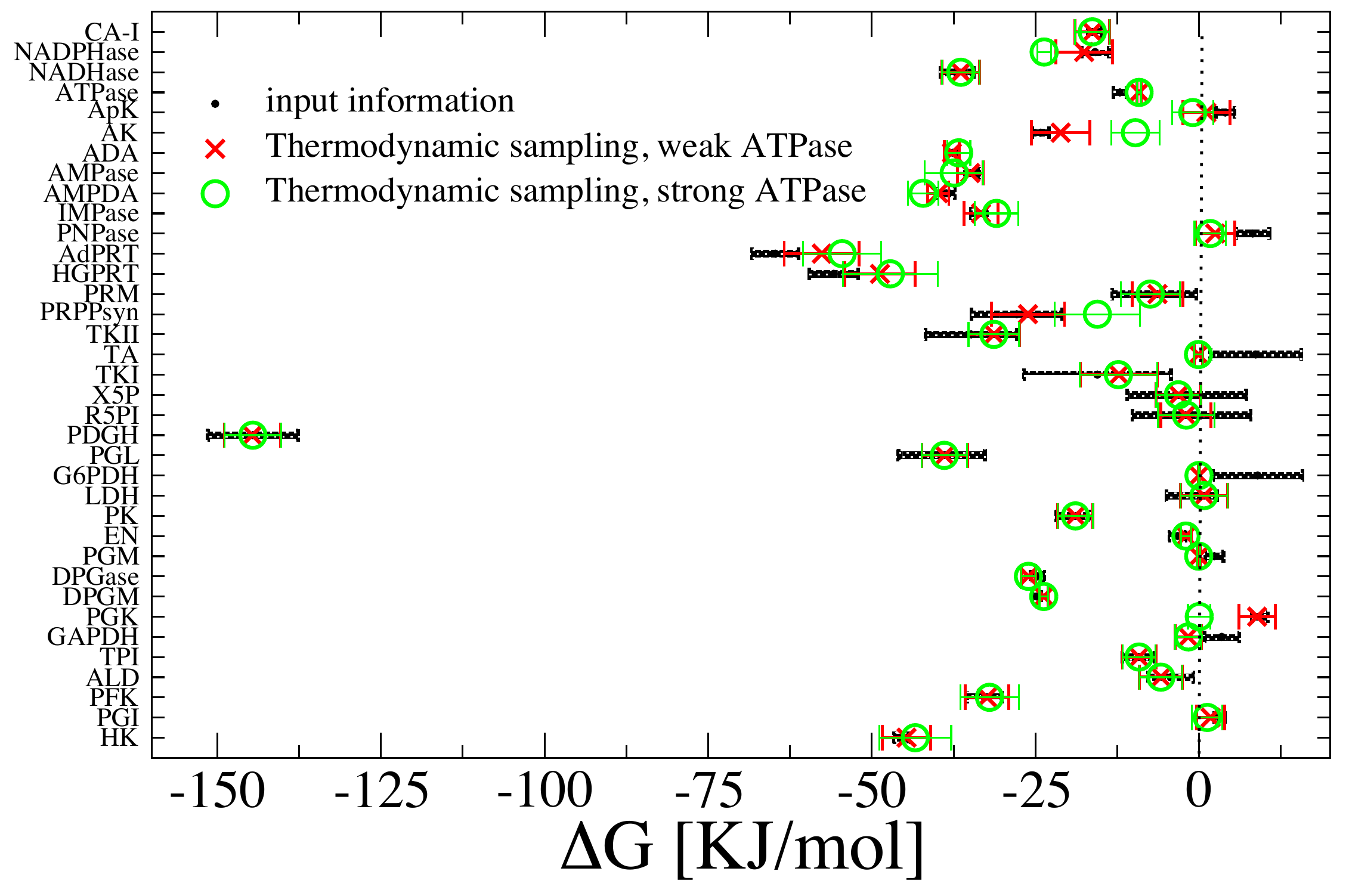}
\includegraphics*[width=0.45\textwidth]{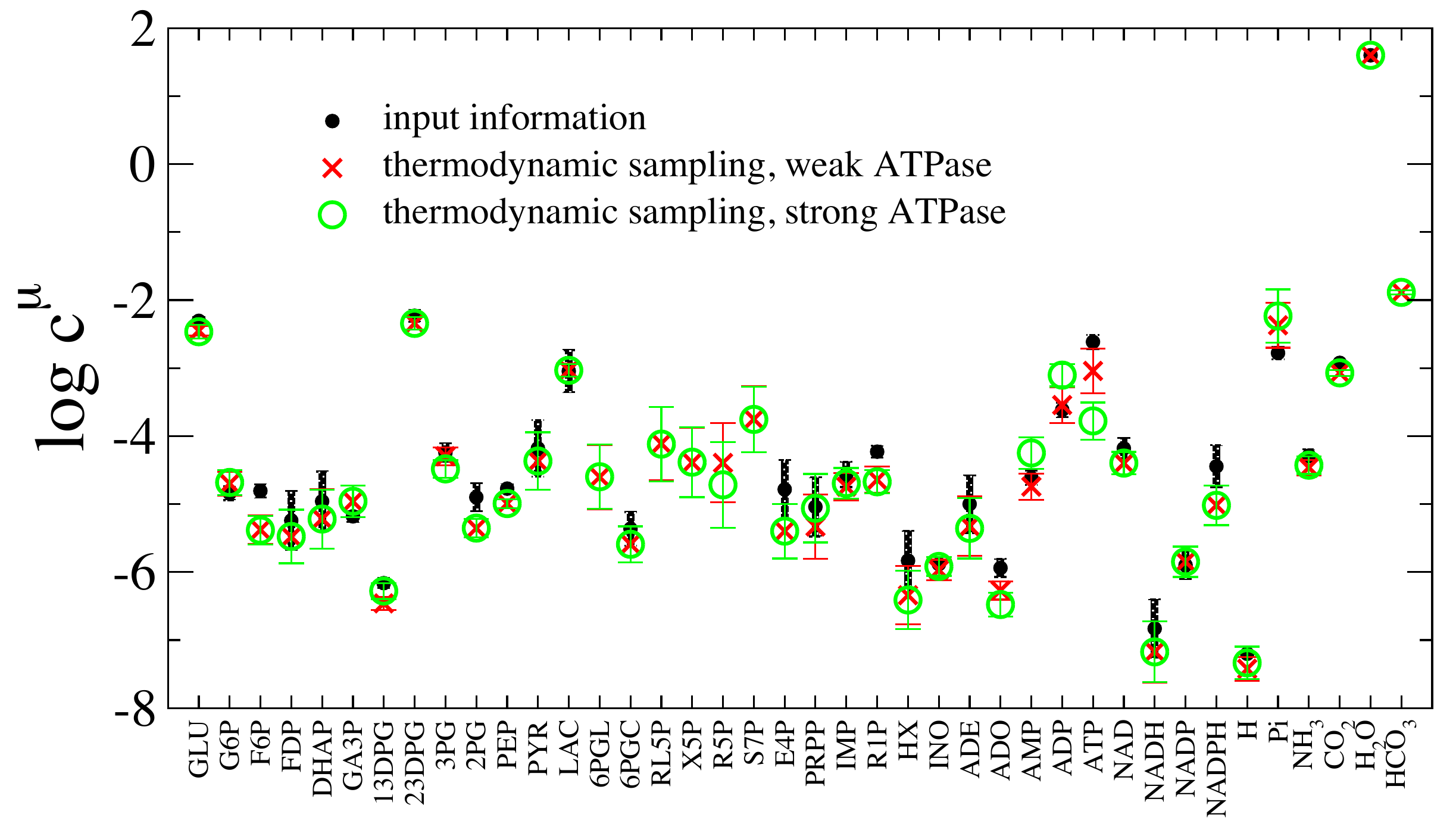}
\caption{Predictions obtained for the hRBC network by Method (b) with rich input information. Top: free energy changes: input information with error (black markers) and values obtained for MBE (red markers) and VNC (green markers) direction assignments. Bottom: measured metabolite (log-)concentrations: input information with error (black markers) and values obtained for MBE (red markers) and VNC (green markers) direction assignments.}
\label{fig3}
\end{center}
\end{figure}

\begin{figure}
\begin{center}
\includegraphics*[width=.45\textwidth]{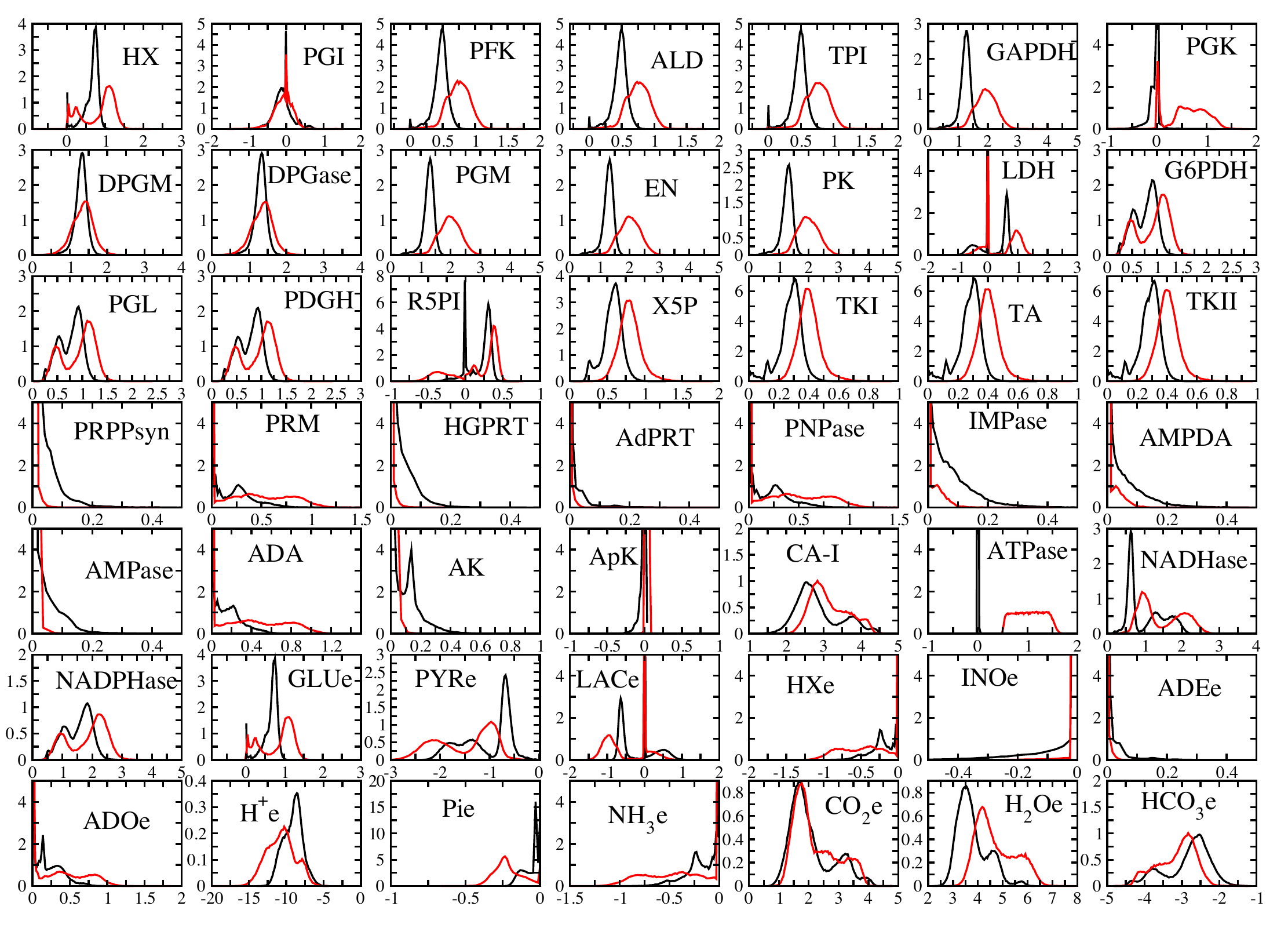}
\caption{Marginal distributions of fluxes for the hRBC network obtained by Method (b) with rich input information under strong (red lines) and weak (black lines) ATPase flux. The reference unit for fluxes is fixed to the measured value of GLC uptake, namely $4 \cdot 10^{-6} \text{mol/s}$.}
\label{fig4}
\end{center}
\end{figure}

Results obtained by Method (b) (without a priori assumptions on reversibility) are displayed in Figs. \ref{fig3} and \ref{fig4}. 
Because the ATPase pump turns out to be weakly active across the solution space (as in \cite{andrea}), we have performed a second set of calculations forcing a strong flux through it (comparable with the GLC uptake, as in\cite{palsson}). In the former case, compared to the Monte Carlo sampling of \cite{palsson} two more reactions are found to be effectively bidirectional in agreement with their standard assignment, i.e. PGK and lactate dehydrogenase (LDH). Note that when the former is operating backward, it activates a futile cycle with the Rapoport-Luebering shunt, a behavior that has been experimentally observed in acidic conditions \cite{futile}. When ATPase flux is large, instead, PGK is constrained in the forward direction. On the other hand, reverse LDH implies a LAC intake and a higher flux of NADHase. Simple biochemical arguments (see Section S5) suggest that the physiologically relevant scenario for hRBCs is that where the ATPase flux is much smaller than the GLC uptake. We shall henceforth focus on this case (corresponding results for strong ATPase flux are reported in Sec. S4).

\begin{figure}
\begin{center}
\includegraphics*[width=0.49\textwidth,angle=0]{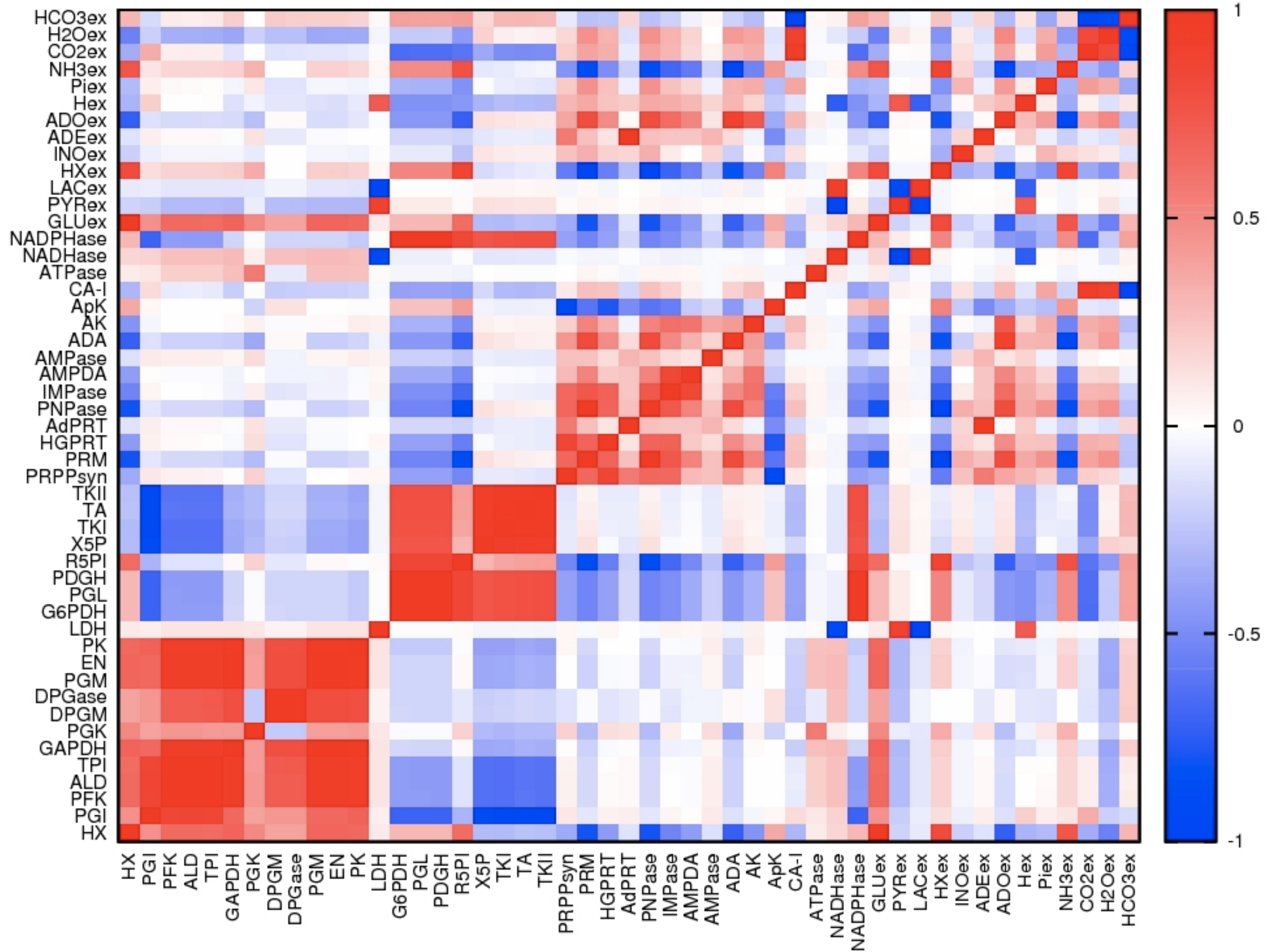}
\caption{Pearson coefficients of the fluxes obtained by Method (b) with rich input information under weak ATPase flux.}
\label{figcorr}
\end{center}
\end{figure}

The pairwise correlations among fluxes (Fig. \ref{figcorr}) show that
glycolysis and the pentose-phosphate pathway form tight modules that
are weakly anti-correlated with each other, mostly through the
reversible phosphoglucoisomerase (PGI) reaction. 
The forward (resp. backward) direction of PGI indeed indicates that glucose is
processed preferentially through glycolysis (resp. pentose phosphate
pathway). Glycolysis can furthermore be broken down into two separated
blocks with a weaker (positive) cross-correlation. The nucleotides
salvage pathway forms instead a weakly correlated module. The network
ultimately presents six bidirectional reactions (PGI, PGK, LDH, R5PI,
ApK, LACe).  Comparing Figs. \ref{fig4} and \ref{figcorr} one sees
that the directions of PGK, LDH and PGI govern the state of ATPase,
NADHase and NADPHase, respectively. For instance, the negative
(resp. positive) part of the LDH distribution corresponds to the high
(resp. low) flux state of the NADHase pump (LDH and NADHase are
strongly anticorrelated).

\subsection{Chemical potential correlations}

\begin{figure}
\begin{center}
\includegraphics*[width=0.45\textwidth,angle=0]{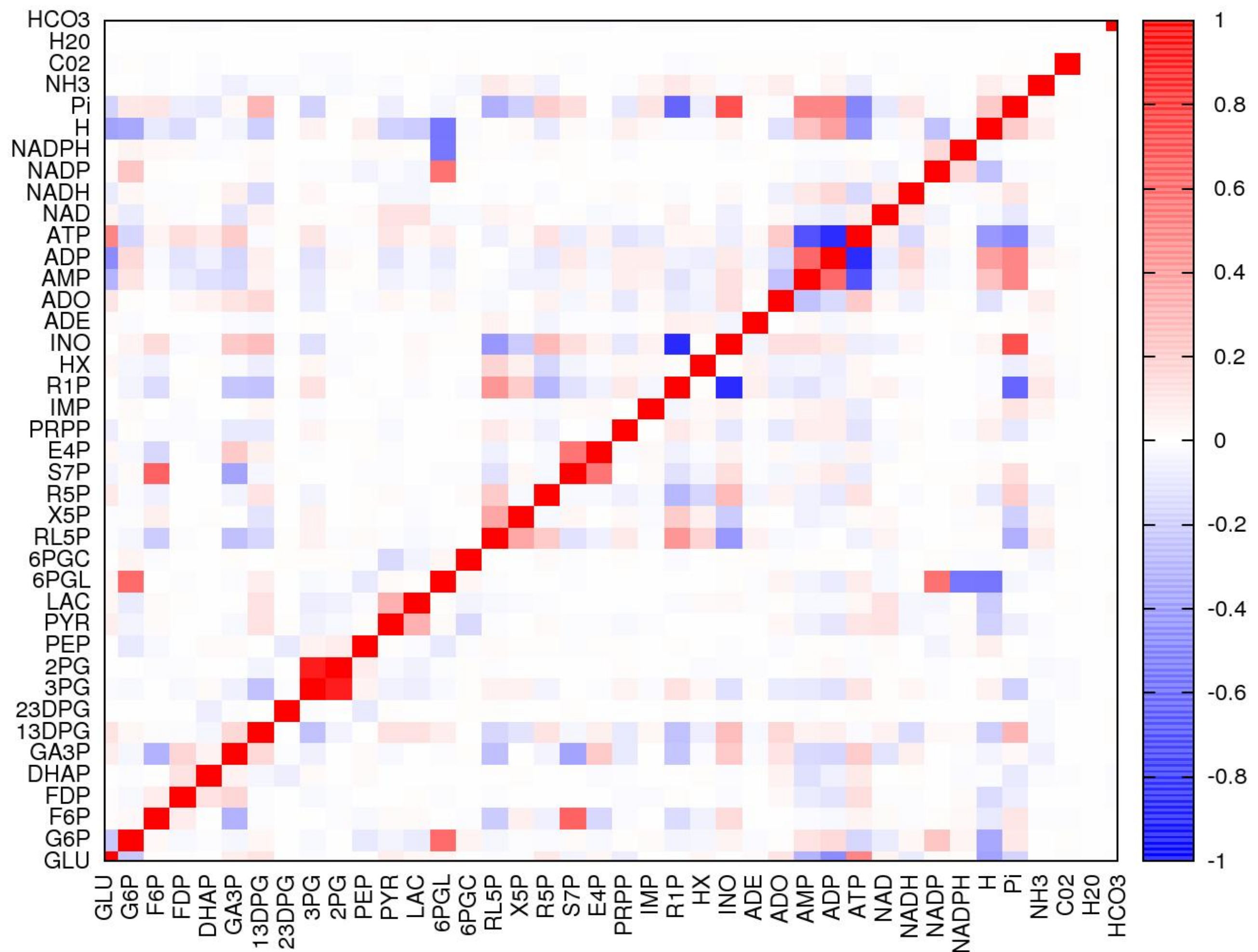}
\caption{Pearson coefficients of the chemical potentials obtained by Method (b) with rich input information under weak ATPase flux.}
\label{Pearson_coeff}
\end{center}
\end{figure}

We hinted above that the coupling of (\ref{vn}) and (\ref{tc})
generates non trivial correlations between the chemical potentials in
feasible states in both Methods (a) and (b). A quantitative look at
this aspect is reported in Fig. \ref{Pearson_coeff}. 
Note that correlations were initially absent, since $P_0(\mathbf{g})$ is assumed
to be a product of uncorrelated independent distributions. The
algorithmic origin of such interdependecies can be understood
considering that chemical potentials are updated dynamically through a
series of reinforcement steps of the form $\alpha s_i \xi^\mu_i$. It
follows that the $g^\mu$'s can ultimately be written as
$g^\mu=g^\mu_{\text{tr}}+\alpha\sum_i k_i \xi^\mu_i$, where
$g_{\text{tr}}^\mu$ is the trial chemical potential sampled from $P_0$
and $k_i$ is an index which is updated (increased or decreased by one
according to the sign of the reaction) each time reaction $i$ tries to
invert. The covariance between chemical potentials can thus be
decomposed as
\begin{equation}\label{correlations}
\langle g^\mu g^\nu \rangle_c=\delta_{\mu\nu}\sigma^2_\mu+\alpha\sum_{i=1}^N 
\xi_i^\mu\langle g_{\text{tr}}^\nu k_i\rangle_c+\alpha\sum_{i=1}^N \xi_i^\nu
\langle g_{\text{tr}}^\mu k_i\rangle_c+\alpha^2\sum_{i,j}\xi_i^\mu 
\xi_j^\nu\langle k_i k_j \rangle_c
\end{equation}
where $\langle . \rangle_c$ stands for the connected correlation, $\sigma_\mu^2$ is the variance of $P_0^\mu$, so that $\langle g_{\text{tr}}^\mu g_{\text{tr}}^\nu \rangle_c=\delta_{\mu\nu}\sigma^2_\mu$. Neglecting correlations between $k_i$ and $k_j$ for $i\neq j$ and between $g_{\text{tr}}^\mu$ and $k_i$ (for each $\mu$ and $i$) this reduces to
\begin{equation}\label{appro_correlations}
\langle g^\mu g^\nu \rangle_c\simeq \alpha^2\sum_{i=1}^N\xi_i^\mu \xi_i^\nu\sigma^2_{k_i}~~~~~(\mu\neq\nu)\;.
\end{equation}
This simply tells us that the dynamics tends to correlate (resp. anti-correlate) metabolites typically appearing on the same (resp. opposite) side of the reaction equations (such as ADP and ATP).

In this respect, it is worthwhile to compute the amount of information gained (or lost) through the sampling algorithms essentially via the build-up of correlations. Note that indeed, since input distributions can be dynamically broadened, a loss of information is actually possible. The rationale behind this flexibility is that we are interested in estimating the ranges of variability of {\it single cell} states, which may exceed the uncertainty derived from experimental measures that are typically performed by averaging over a large number of cells. Now the information gain is related to the difference in entropy of the initial and final distributions, which is particularly simple to estimate assuming to be dealing with Gaussian distributions. Indeed the entropy of a Gaussian $Q(\mathbf{g})$ is given by \cite{entro}
\begin{equation}
S[Q(\textbf{g})]=\frac{1}{2}\text{tr}[\log(\textbf{K})]+\frac{M}{2}[1+\log(2\pi)]
\end{equation}
where $\textbf{K}$ is the covariance matrix of the $g^\mu$'s and $M$ is the dimensionality of the space, i.e. the number of metabolites. It follows that
\begin{equation}
I =-(S[P(\textbf{g})]-S[P_0(\textbf{g})])=\frac{1}{2}\sum_{\mu=1}^M[\log(\lambda^\mu_0)-\log(\lambda^\mu)]\;,
\end{equation}
where $\lambda_0^\mu$ and $\lambda^\mu$ are respectively the eigenvalues of the covariance matrix for the initial and final distributions, is the information gain. Surprisingly, the information gain thus computed is found to be negative ($I\simeq -3.98$) for Method (a), and positive ($I\simeq 4.12$) for Method (b), indicating that the thermodynamic sampling algorithm allows for a refinement of the input information. The rank plot of eigenvalues (see Fig. S5) shows clearly that Method (a) is capable of extracting information only from the metabolites with largest $\mathbf{K}$ eigenvalues (associated with highly uncertain concentrations) and is generically outperformed by Method (b).

\begin{figure}
\begin{center}
\includegraphics*[width=0.49\textwidth]{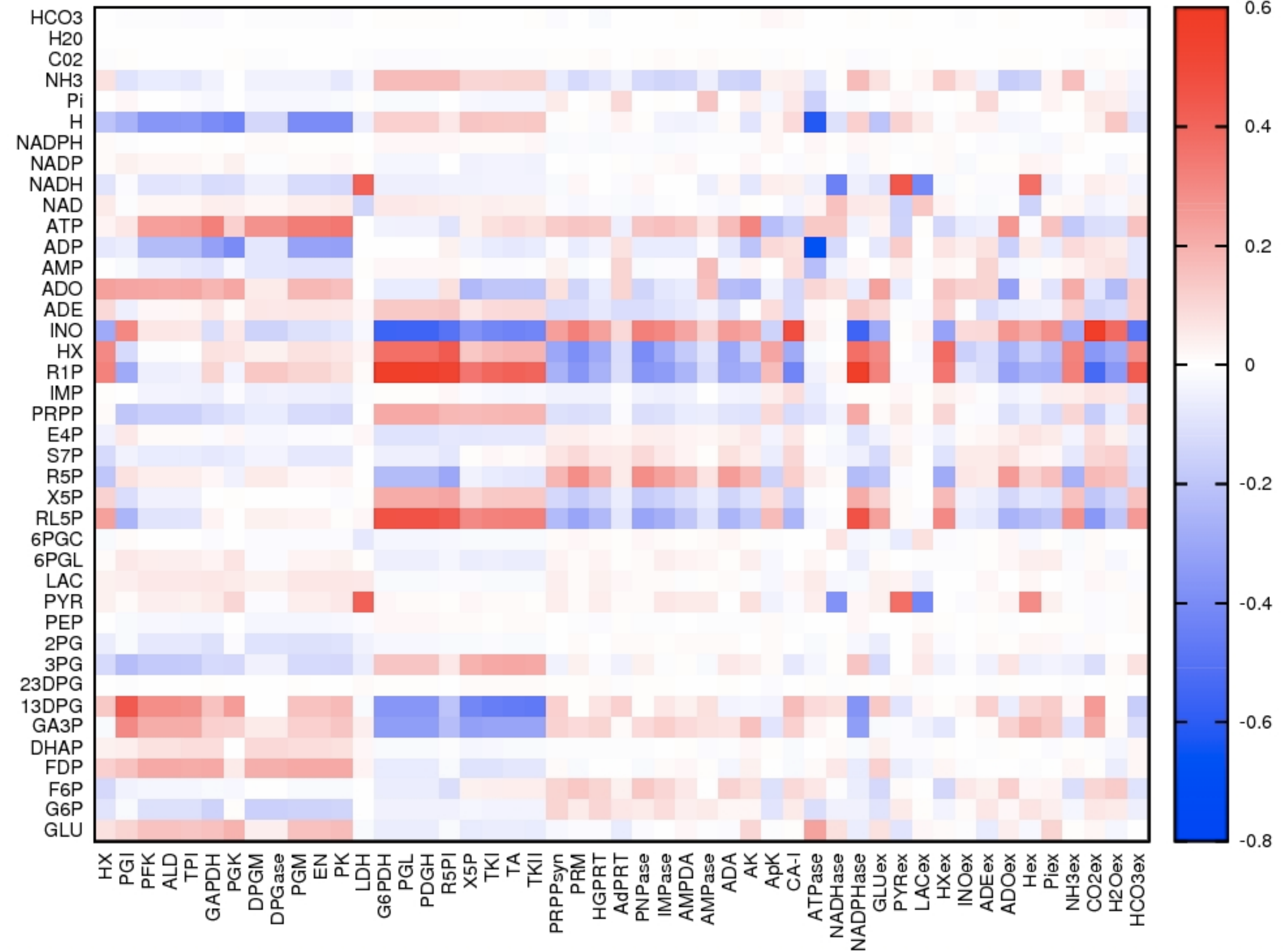}
\caption{Correlations between chemical potentials and fluxes (see (\ref{chi})) obtained by Method (b) with rich input information under weak ATPase flux.}
\label{eta2}
\end{center}
\end{figure}

In order to spot key-metabolites whose concentration variability bears a strong influence on the global organization of fluxes, we can quantify the correlations between fluxes and concentrations emerging from Method (b) by computing the index
\begin{equation}\label{chi}
\chi_i^\mu=\frac{\left\langle (\nu_i-\left\langle\nu_i\right\rangle)(g^\mu-\left\langle g^\mu\right\rangle)\right\rangle}{\sqrt{\left\langle(\nu_i-\left\langle \nu_i\right\rangle)^2\right\rangle\left\langle(g^\mu-\left\langle g^\mu\right\rangle)^2\right\rangle}}\;~,
\end{equation}
which correlates the fluctuations in chemical potentials to those in fluxes. This matrix is shown in Fig. \ref{eta2}. 
High concentrations of (1,3)-DPG and GA3P appear to favor the glycolytic pathway over the pentose-phosphate pathway, whereas the nucleotide rescue pathways is activated by high levels of INO and R5P and low levels of HX and R1P, whose concentration is instead correlated with high flux through the pentose-phosphate group. Note that concentrations of metabolites only appearing in far from equilibrium reactions (like (2,3)-DPG) are seldom modified by the algorithm and thus appear to be particularly stable, as the corresponding $\chi_i^\mu$ are weakly dependent on fluxes.

\subsection{Chemical energy balance}

The network nodes that are more tightly connected to the biological functionality of the haematids are the three $ase$ pumps: ATPase (that regulates the internal osmotic pressure and the transmembrane potential), NADHase (that reduces hemoglobin from its metastable state) and NADPHase (that maintains the redox state of the cell by reducing glutathione). Their respective work loads can be estimated from basic biochemical parameters.

The work carried out by the ATPase pump can be written as
\begin{equation}
W_{\text{ATP}} = - FV_0 + RT \left[ 3 \log 
\frac{c_{\text{Na}^+_{(ext)}}}{c_{\text{Na}^+_{(in)}}}+
2\log \frac{c_{\text{K}^+_{(in)}}}{c_{\text{K}^+_{(ext)}}} \right]\;,
\end{equation}
where $F \simeq 96~ \text{KJ/(mol V)}$ is the Faraday constant and $V_0\simeq -12~\text{mV}$ is the transmembrane potential, while $c_{\text{Na}^+_{(ext)}}/c_{\text{Na}^+_{(in)}}\simeq 20$ and $c_{\text{K}^+_{(in)}}/c_{\text{K}^+_{(ext)}}\simeq18$ are the ratios between external and internal levels of sodium and potassium ions. At $T=310 \text{K}$, we obtain $W_{\text{ATP}} \simeq 50 \text{ KJ/mol}$.

The work carried out by the NADHase pump, specifically by cytochrome b5 reductase, can instead be estimated from the standard values of the redox potential of the couple Fe$^{2+}/$Fe$^{3+}$ in the heme group of hemoglobin (namely $V_1 \simeq 60 \text{ mV}$ \cite{leninger}) upon assuming normal levels of meta-hemoglobin ($2\%$ of the total). We obtain
\begin{equation} 
W_{\text{NADH}} = -2 F V_1 + 2RT \log \frac{c_{\text{Hb}}}{c_{\text{MetHb}}} \simeq 10 \text{ KJ/mol}\;.
\end{equation}
(The factor two comes from the fact that the reductase enzyme couples
each NADHase with the reduction of two molecules of oxidized
hemoglobin.)

Finally we can estimate the work done by the NADPHase pump,
glutathione reductase, from the standard redox potential of the pair
2GSH/GSSG, i.e. $V_2 \simeq -230 mV$, and from their concentrations
measured in human red cells, given respectively by $ c_{\text{GSH}}
\simeq 3.2 \cdot 10^{-3} \text{ M} $, $ c_{\text{GSSG}} \simeq 6.5
\cdot 10^{-5} \text{ M}$ \cite{leninger}.  We find
\begin{equation}
W_{\text{NADPH}} = - FV_2 + RT \log\frac{c_{\text{GSH}}^2}{c_{\text{GSSG}}}  \simeq 55  \text{ KJ/mol}.
\end{equation}

The total amount of work per unit time performed by these three
processes can now be written as
\begin{equation}\label{wdot}
\dot{W}= \nu_{\text{ATPase}} W_{\text{ATP}} + \nu_{\text{NADHase}} 
W_{\text{NADH}} + \nu_{\text{NADPHase}} W_{\text{NADPH}}\;,
\end{equation}
where $\nu_{\text{ATPase}}$ is the flux of ATPase (and similarly for
the remaining pumps). At the same time, the entropy produced by the
cell per unit time reads
\begin{equation}
T\dot{S} \equiv - \sum_{i\in R\setminus U} \nu_i \Delta G_i -
\sum_{\mu \in U} u^\mu \cdot (g_{ext}^\mu-g^\mu)-\dot{W}\; ,
\end{equation}
where the first sum coincides with the quadratic form defined by the
stoichiometric matrix of intracellular reactions, i.e. $\sum_{i,\mu}
\nu_i \xi_i^\mu g^\mu$, and the second is over cross-membrane
transport processes (uptakes), with $u^\mu$ the uptake flux of
metabolite $\mu$. In order to obtain a realistic description of the
processes, after normalizing our computed fluxes with respect to the
average value of the GLC uptake ($u \simeq 4 \cdot 10^{-6} \text{
  mol/(l s)}$ \cite{tornbum}), we scaled them by the cell volume
($V_{\text{hRBC}}\simeq 90\text{ fl}$). The na\"ive thermodynamic
efficiency of the hRBC can finally be evaluated as
\begin{equation}
\eta =\frac{\dot{W}}{\dot{W}+T\dot{S}}
\end{equation}
In Fig. \ref{fig5} we display the distributions of $\dot{W}$,
$T\dot{S}$, $\dot{U} =\dot{W} + T\dot{S}$ and $\eta$ obtained from
Method (b) with weak and strong ATPase flux. 
\begin{figure}
\begin{center}
\includegraphics*[width=.49\textwidth]{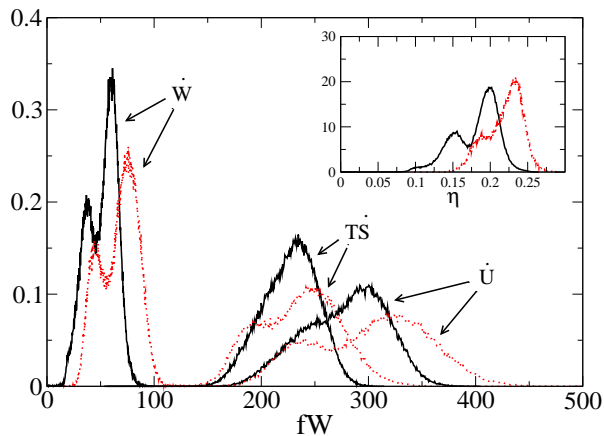}
\caption{Distributions of work done ($\dot{W}$), entropy produced ($T\dot{S}$) and energy flow 
($\dot{U}$) per unit time for the hRBC metabolism with strong (dotted line) and weak (straight line) ATPase flux. Inset: distribution of efficiency ($\eta$) in the same conditions.}
\label{fig5}
\end{center}
\end{figure}
The two peaks in the distributions mirror those appearing in the flux distributions of the
NADH and NADPH pumps. Remarkably, the computed efficiency of the hRBC
is not far from that corresponding to optimal microbial growth, which
was estimated to be close to 24\% \cite{westeroff}. Note that the work performed per unit time by the NADPHase pump dominates the sum (\ref{wdot}). Therefore the two peaks appearing in the distribution of $\dot W$ mirror those appearing in the distribution of $\nu_{\text{NADPHase}}$.
 
\section{Conclusion}

Constraint-based models of cellular metabolism are important tools to
analyze and predict the chemical activity and response to
perturbations of cells without relying on kinetic and transport
details that are often unavailable. In such frameworks, assessing the
metabolic capabilities of a cell requires the exploration of a high
dimensional space representing flux and chemical potential
configurations compatible with mass- and energy-balance
constraints. The complexity of the ensuing problem can in some cases
be reduced by applying reasonable (though necessarily ad hoc)
optimality criteria. Most often, however, one needs to sample
``physiological configurations'' from the solution space starting from
a priori biochemical knowledge that could be noisy. The methods
proposed here are designed to retrieve distributions of fluxes and
chemical potentials (or concentrations) essentially by exposing and
building up correlations between variables. This is substantially
different from other approaches (e.g. \cite{hoppe,kummel}), where
either the flux distributions compatible with given chemical
potentials or the reverse are sought. Disposing of reliable prior
biochemical information is obviously a limiting factor in our case as
well. The advantage lies in the fact that working with correlations
allows for a greater flexibility in treating the biochemical input
data. Besides providing information on feasible physiological
concentration ranges, flux distributions and reaction directionality,
thermodynamic sampling can be employed to evaluate the responsiveness
of fluxes to fluctuations in concentrations (or the reverse) and to
assess the thermodynamic efficiency of a cell. In the case considered
here (the hRBC, where a comparison with results obtained by sampling
the solution space of mass-balance equations is possible), we have
estimated $\eta\simeq 0.24$ assuming that the functional core of its
metabolism lies in the three pumps (ATPase, NADHase, NADPHase). A
similar calculation can be carried out on more complex genome-scale
systems (e.g. E. coli) for a modest increase of computational costs
(work in progress).

\acknowledgements This work is supported by the DREAM Seed Project of the Italian Institute of Technology (IIT) and by the joint IIT/Sapienza Lab ``Nanomedicine''.

\appendix

\section{Role of thermodynamic constraints: a simple example}

\begin{figure}
\begin{center}
\includegraphics*[width=0.48\textwidth]{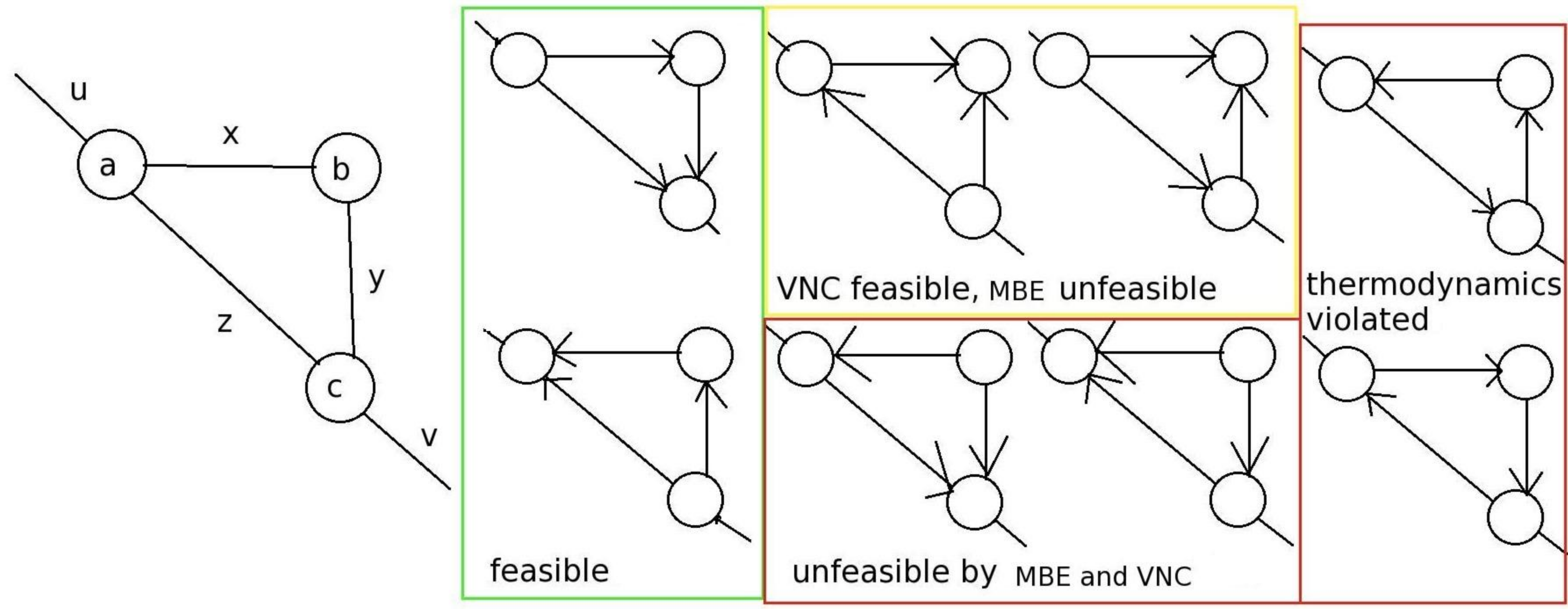}
\caption{\label{fluxis}Network module with $3$ internal reactions,
  $x$, $y$ and $z$ between $3$ chemical species $a$, $b$ and $c$ and
  $2$ uptake fluxes $u$ and $v$.  Unit stoichiometry is assumed.}
\end{center}
\end{figure}

In order to clarify how thermodynamic constraints translate into
bounds on the degrees of freedom, consider the small module shown in
Fig. \ref{fluxis}. A priori, it admits $2^5 = 32$ different direction
assignments, corresponding to the 8 possible states for the 3 internal
fluxes $x$, $y$ and $z$, times the 4 possible states of the boundary
fluxes $u$ and $v$.
Assuming unit stoichiometry, one easily sees that 6 of the 8 states
allowed for $(x,y,z)$ are thermodynamically feasible, corresponding to
the number of ways in which one can order the chemical potentials of
metabolites $a$, $b$ and $c$. The excluded configurations present {\em
  unfeasible cycles}. Mass constraints further reduce the space of
possible directions. Mass-balance equations (MBE) impose that each
metabolite should be produced and consumed by at least one reaction,
leaving only two feasible direction assignments. The softer Von
Neumann conditions (VNC) instead force each metabolite to be produced
at least by one reaction, leaving four possible states. Once the
possible directions of the boundary fluxes $u$ and $v$ are considered,
one ends up with two thermodynamically and stoichiometrically feasible
configurations for MBE and eight for VNC (down from 32).

Now let us focus on a specific flux model (say MBE) and note that it
enforces $u=v$, $x=y$ and $v=x+z$. This leaves two free parameters,
e.g. $v$ and $x-z=\lambda$, and it is easy to show that the exclusion
of unfeasible cycles implies $|\lambda|\leq u$. Similar though more
lengthy arguments can be formulated for the Von Neumann flux model
VNC.

Further examples are discussed e.g. in \cite{beard1,book}.

\section{hRBC data}

\begin{figure}
\begin{center}
\includegraphics*[width=0.48\textwidth]{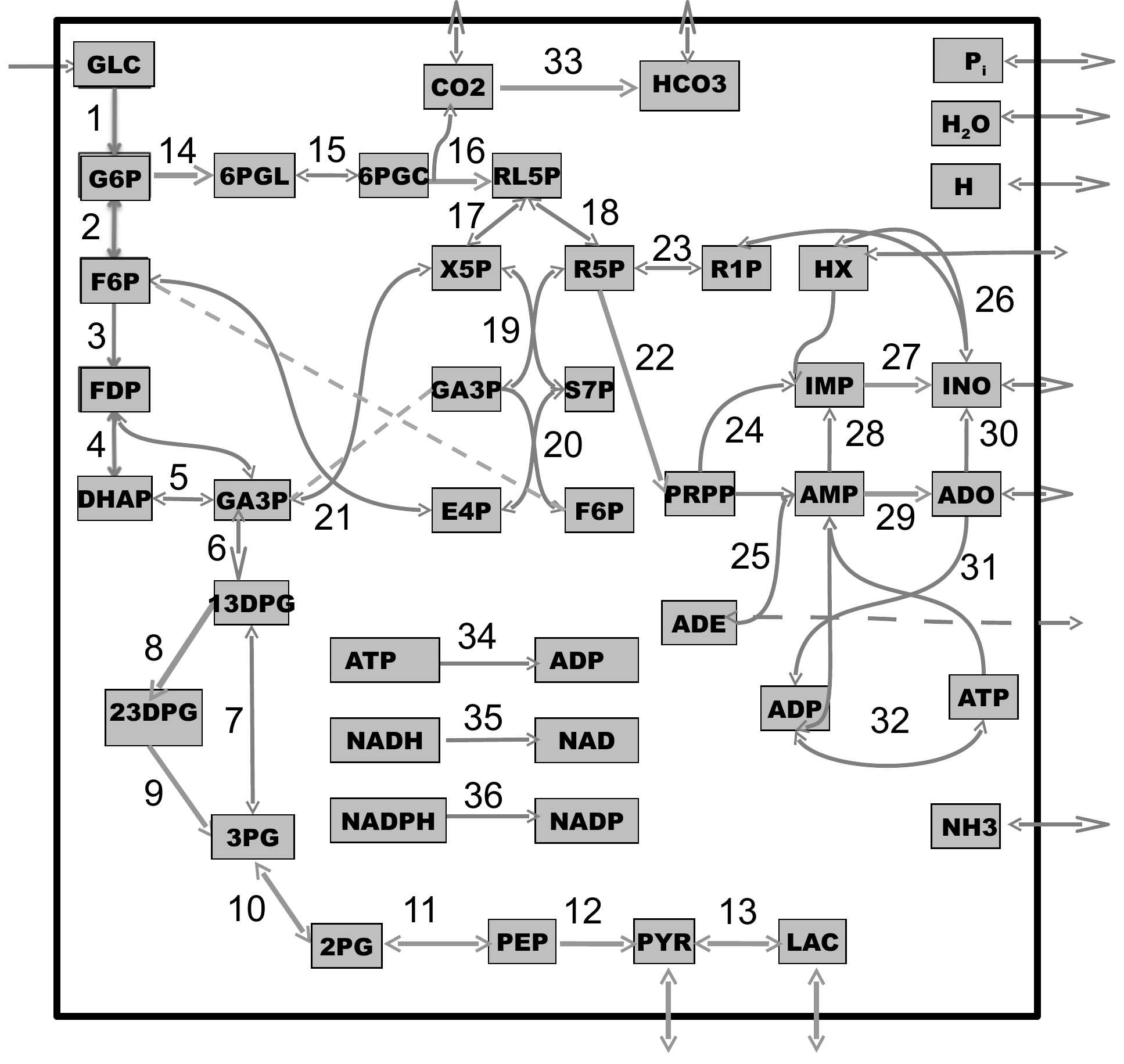}
\caption{The hRBC network employed in this study. }  
\label{network}
\end{center}
\end{figure}

The hRBC metabolic network employed for this study is shown in Fig. \ref{network}.
It is formed by $40$ metabolites (listed in Table \ref{metab})  interacting through $35$ intracellular reactions (listed in Table \ref{react}) and subject to $12$ (for Method (a)) or 13 (for Method (b)) uptakes.  (Bicarbonate outtake was only accounted for in Method (b).) The network can be divided in three main pathways, namely glycolysis (reactions 1-13), the pentose phosphate pathway (reactions 14-21), and a nucleotide salvage pathway (reactions 22-32). The directions displayed correspond to the standard physiological assignment. The network coincides with the reconstruction presented in \cite{palsson}, except for the inclusion of bicarbonate (HCO$_3^-$), of the  carbonic anhydrase reaction (33) and of a bicarbonate uptake. Table \ref{metab} also provides the standard chemical potential for each metabolite and an estimate of the intracellular concentration (when available). Potentials are computed in the standard biochemical state, i.e. in acqueous solution at fixed temperature, pressure, ionic strength and pH and are available in \cite{kummel}, where they are calculated according to the prescriptions of \cite{alberty,alberty2} at $T=298$ K, $P=1$ atm, pH $=7.6$ and ionic strenght I $=0.15$ M. The physiological conditions of the hRBC are known to be slightly different ($T=310$ K, pH $=7.2$), but such differences do not affect significantly the results of our procedures. The estimated intracellular concentration ranges are obtained from measurements in different settings. It is worth to notice that the experimental errors on such values of concentrations reflect an essential uncertainty  rather than statistical fluctuations from cell to cell, since standard measurements of concentrations are usually carried out averaging over many ($10^2 - 10^8$) cells. 
\begin{table*}
\begin{center}
\footnotesize{
\begin{tabular}{| c | l | c | l | }
\hline 
{\scriptsize Abbr.} & Compound name & $g_0$ [KJ/mol] & $c$ [M]\\ 
\hline
GLC$^*$ & {\scriptsize Glucose} & $-387$ & $5\pm1 \cdot 10^{-3}$ \cite{beutler}\\
G6P & {\scriptsize Glucose-6-phosphate} & $-1281$ & $4\pm1 \cdot 10^{-5}$ \cite{beutler}\\
F6P & {\scriptsize Fructose-6-phosphate} & $-1278$ & $1.3\pm0.5 \cdot 10^{-5}$ \cite{beutler}\\
FDP & {\scriptsize Fructose-1,6-diphosphate} & $-2171$ & $2.7\pm1 \cdot 10^{-6}$ \cite{beutler}\\
DHAP & {\scriptsize Dihydorxyacetone phosphate} & $-1070$ & $1.7\pm0.1 \cdot 10^{-5}$ \cite{minakami}\\
GA3P & {\scriptsize Glyceraldehyde-3-phosphate} & $-1078$ & $5.7\pm1 \cdot 10^{-6}$ \cite{minakami}\\
13DPG & {\scriptsize 1,3-Diphosphoglycerate} & $-2191$ & $1\pm0.5 \cdot 10^{-6}$ \cite{garrett}\\
23DPG & {\scriptsize 2,3-Diphosphoglycerate} & $-2240$ & $4\pm3 \cdot 10^{-3}$ \cite{garrett}\\
3PG & {\scriptsize 3-Phosphoglycerate} & $-1332$ & $4\pm2 \cdot 10^{-5}$ \cite{minakami}\\
2PG & {\scriptsize 2-Phosphoglycerate} & $-1326$ & $1.4\pm0.5 \cdot 10^{-5}$ \cite{minakami}\\
PEP & {\scriptsize Phosphoenolpyruvate} & $-1181$ & $1.7\pm0.2 \cdot 10^{-5}$ \cite{beutler}\\
PYR$^*$ & {\scriptsize Pyruvate} & $-341$ & $8\pm6 \cdot 10^{-5}$ \cite{minakami}\\
LAC$^*$ & {\scriptsize Lactate} & $-297$ & $1.4\pm0.5 \cdot 10^{-3}$ \cite{beutler}\\
6PGL & {\scriptsize 6-Phosphogluco-lactone} & $-1352$ &  \\
6PGC & {\scriptsize 6-Phosphogluconate} & $-1353$ &  $5\pm2 \cdot 10^{-6}$ \cite{kirkman}\\
RL5P & {\scriptsize Ribulose-5-phosphate} & $-1201$ &  \\
X5P & {\scriptsize Xylusose-5-phosphate} & $-1203$ &  \\
R5P & {\scriptsize Ribose-5-phosphate} & $-1202$ & \\
S7P & {\scriptsize Sedoheptulose-7-phosphate} & $-1336$ &  \\
E4P & {\scriptsize Erythrose-4-phosphate} & $-1125$ & $5\pm2 \cdot 10^{-5}$ \cite{magnani} \\
PRPP & {\scriptsize 5-Phosphoribosyl-1-pyrophosphate} & $-2949$ & $5\pm1 \cdot 10^{-5}$  \cite{bionumbers}\\
IMP & {\scriptsize Inosine monophosphate} & $-774$ & $3\pm1 \cdot 10^{-5}$  \cite{tomoda, geibuhler}\\
R1P & {\scriptsize Ribose-1-phosphate} & $-1194$ & $6\pm5 \cdot 10^{-6}$    \\
HX$^*$ & Hypoxanthine & $274$ & $2\pm1 \cdot 10^{-6}$   \cite{bionumbers} \\
INO$^*$ & Inosine & $119$ & $1.3\pm0.5 \cdot 10^{-6}$    \cite{bionumbers}\\
ADE$^*$ & Adenine & $538$ & $1.3\pm0.7 \cdot 10^{-5}$    \cite{bionumbers}\\
ADO$^*$ & Adenosine & $378$ & $1.3\pm0.3 \cdot 10^{-6}$    \cite{bionumbers} \\
AMP & {\scriptsize Adenosine monophosphate} & $-514$ & $8\pm1 \cdot 10^{-4}$    \cite{leninger}\\
ADP & {\scriptsize Adenosine diphosphate} & $-1384$ & $1.0\pm0.1 \cdot 10^{-3}$    \cite{leninger}\\
ATP & {\scriptsize Adenosine triphosphate} & $-2250$ & $7.9\pm0.1 \cdot 10^{-3}$    \cite{leninger}\\
NAD & {\scriptsize Nicotinamide adenine dinucleotide} & $1145$ & $7\pm2 \cdot 10^{-5}$\cite{minakami}    \\
NADH & {\scriptsize Nicotinamide adenine dinucleotide(R)} & $1209$ &  around $10^{-7}$ (ext)  \\
NADP & {\scriptsize Nicotinamide adenine dinucleotide phosphate} & $260$ & $1.4\pm0.5 \cdot 10^{-6}$ \cite{omaci}   \\
NADPH & {\scriptsize Nicotinamide adenine dinucleotide phosphate (R)} & $325$ & $4\pm2 \cdot 10^{-5}$    \cite{omaci}\\
H$^*$ & {\scriptsize Hydrogen ion} & $0$ &  $10^{-7.2}$ \cite{sved}\\
Pi$^*$ & {\scriptsize Inorganic phosphate} & $-1055$ & $1.0\pm0.5 \cdot 10^{-3}$    \cite{leninger}\\
NH$_3^*$ & {\scriptsize Ammonia} & $96$ & $5\pm1 \cdot 10^{-5}$    \cite{oldyale}\\
CO$_2^*$ & {\scriptsize Carbon dioxide} & $-543$ & $1.2\pm0.5 \cdot 10^{-3}$    \cite{mcgilvery}\\
H$_2$O$^*$ & {\scriptsize Water} & $-149$ &  solvent  \\ 
HCO$_3^*$ & {\scriptsize Bicarbonate} & $-710$ & $1.6\pm0.1 \cdot 10^{-2}$    \cite{mcgilvery}\\
\hline
\end{tabular}}
\caption{Metabolites included in the human red blood cell metabolic network used in this study. $g_0$ represents the standard chemical potential of the metabolite, while $c$ denotes an indicative value for the experimentally estimated intracellular concentration, when available. Compounds marked by an asterisk are subject to uptakes. All data were searched through the Bionumbers database \cite{bionumbers}; pointers to explicit references are given when available.}
\label{metab}
\end{center}
\end{table*}
\begin{table*}
\begin{center}
\footnotesize{
\begin{tabular}{| c | c | l | l |}
\hline 
Nr & Abbr & Enzyme & reaction \\ 
\hline
1 & HK & {\scriptsize Hexokinase} & {\scriptsize $GLC+ATP \to G6P+ADP+H$}   \\
2 & PGI & {\scriptsize Phosphoglucoisomerase} & {\scriptsize $G6P \leftrightarrow F6P$} \\
3 & PFK & {\scriptsize Phosphofructokinase} & {\scriptsize $F6P+ATP \to FDP+ADP+H$} \\
4 & ALD & {\scriptsize Aldolase} & {\scriptsize $FDP \leftrightarrow GA3P +DHAP$}  \\
5 & TPI & {\scriptsize Triose phosphate isomerase} & {\scriptsize $DHAP \leftrightarrow GA3P$}  \\
6 & GAPDH & {\scriptsize GLyceraldehyde phosphate dhydrogenase } & {\scriptsize $GA3P+NAD+Pi \leftrightarrow 13DPG+NADH+H$}  \\
7 & PGK & {\scriptsize Phosphoglycerate kinase} & {\scriptsize $13DPG + ADP \leftrightarrow 3PG
+ ATP$} \\
8 & DPGM & {\scriptsize Diphosphoglyceromutase} & {\scriptsize $13DPG  \to 23DPG + H$}  \\
9 & DPGase & {\scriptsize Diphosphoglycerate phosphatase} & {\scriptsize $ 23DPG + H_{2} O \to 3PG + Pi$}  \\
10 & PGM & {\scriptsize Phosphoglyceromutase} & {\scriptsize $ 3PG \leftrightarrow 2PG $} \\
11 & EN & {\scriptsize Enolase} & {\scriptsize $2PG   \leftrightarrow PEP+H_{2}O$} \\
12 & PK & {\scriptsize Pyruvate kinase} & {\scriptsize $PEP+ADP+H \to PYR + ATP$}\\
13 & LDH & {\scriptsize Lactate dehydrogenase} & {\scriptsize$ PYR +NADH +H \leftrightarrow LAC+NAD$}\\
14 & G6PDH & {\scriptsize Glucose-6-phosphate dehydrogenase} & {\scriptsize$G6P +NADP \to 6PGL+NADPH+H$} \\
15 & PGL & {\scriptsize 6-phosphoglyconolactonase} & {\scriptsize$6PGL +H_2O \leftrightarrow 6PGC+H$} \\
16 & PDGH & {\scriptsize 6-phosphoglycoconate dehydrogenase} &{\scriptsize $6PGC + NADP \to RL5P+NADPH +CO_2$}\\
17 & R5PI & {\scriptsize Ribose-5-phosphate isomerase} &{\scriptsize $RL5P \leftrightarrow R5P$} \\
18 & X5P & {\scriptsize Xylulose-5-phosphate
epimerase} &{\scriptsize $RL5P \leftrightarrow X5P$}\\
19 & TKI & {\scriptsize Transketolase I} &{\scriptsize $X5P + R5P \leftrightarrow S7P + GA3P$} \\
20 & TA & {\scriptsize Transaldolase} &{\scriptsize $GA3P +S7P \leftrightarrow E4P +F6P$} \\
21 & TKII & {\scriptsize Transketolase} & {\scriptsize$X5P+E4P \leftrightarrow F6P+GA3P$} \\
22 & PRPPsyn &{\scriptsize Phosphoribosyl
pyrophosphate synthetase} &{\scriptsize $R5P+ATP \to PRPP+AMP$} \\
23 & PRM &{\scriptsize Phosphoribomutase} &{\scriptsize $R1P \leftrightarrow R5P$} \\
24 & HGPRT &{\scriptsize Hypoxanthine guanine phosphoryl transferase} &{\scriptsize $PRPP+HX+H_2O \to IMP +2Pi$} \\
25 & AdPRT &{\scriptsize Adenine phosphoribosyl transferase} &{\scriptsize $PRPP+ADE+H_2O \to AMP +2Pi$} \\
26 & PNPase &{\scriptsize Purine nucleoside phosphorylase} &{\scriptsize $INO+Pi \leftrightarrow HX +R1P$} \\
27 & IMPase &{\scriptsize Inosine monophosphatase} &{\scriptsize $IMP +H_2O \to INO +Pi +H$} \\
28 & AMPDA &{\scriptsize Adenosine monophosphate
  deaminase} &{\scriptsize $AMP +H_2O \to IMP+NH_3$} \\
29 & AMPase &{\scriptsize Adenosine monophosphate
 phosphohydrolase} & {\scriptsize$AMP+H_2O \to ADO+Pi+H$} \\
30 & ADA &{\scriptsize Adenosine deaminase} &{\scriptsize $ADO + H_2O \to INO+NH_3$}\\
31 & AK &{\scriptsize Adenosine kinase} & {\scriptsize$ADO+ATP \to ADP+AMP$}\\
32 & ApK &{\scriptsize Adenylate kinase} &{\scriptsize $2ADP \leftrightarrow ATP+AMP$}\\
33 & CA-I &{\scriptsize Carbonic anhydrase} &{\scriptsize $CO_2 +H_2O \leftrightarrow HCO_3 + H$}\\
34 & ATPase & {\scriptsize Sodium-Potassium ionic pump} & {\scriptsize $ATP + H_2 O \to ADP + Pi $}  \\ 
35 & NADHase & {\scriptsize Cytochrome-b5 reductase} &{\scriptsize $NADH \to NAD + H $}   \\ 
36 & NADPHase & {\scriptsize Glutathion reductase} & {\scriptsize $NADPH  \to NADP + H $}  \\
\hline
\end{tabular}}
\caption{Intracellular reactions included in the human red blood cell metabolic network used in this study.}
\label{react}
\end{center}
\end{table*}

Finally, in Table \ref{ext} we report the reference values for blood tests \cite{wiki} of the level of metabolites of interest in serum. 
\begin{table}
\begin{center}
\footnotesize{
\begin{tabular}{| c |c | }  
\hline 
Compound & Range (M) \\ 
\hline
 GLC & $ 4-6 \cdot 10^{-3}$ \\
 PYR & $ 3-10 \cdot 10^{-5}$  \\
 LAC &  $ 0.5-2.2 \cdot 10^{-3}$ \\
 H$^+$ & $ 10^{-7.3} - 10^{-7.45} $\\
  Pi & $0.8-1.5 \cdot 10^{-3}$\\
 NH$_3$ & $ 1-6 \cdot 10^{-5} $\\
 CO$_2$ & $2-3 \cdot 10^{-2}$\\
 HCO$_3^-$ & $1.8-2.3 \cdot 10^{-2}$\\
\hline
\end{tabular}}
\caption{Estimated concentration levels in blood serum.}
\label{ext}
\end{center}
\end{table}

\section{Details on the implementation of Methods (a) and (b)}

Here we discuss in some detail four aspects connected to the implementation of Methods (a) and (b), namely
\begin{itemize}
\item the minimal amount of input information on chemical potentials (to be included in $P_0(\mathbf{g})$) needed to reconstruct the free energy landscape via Method (a);
\item the form of the input information on chemical potentials (i.e. of $P_0(\mathbf{g})$) in the poor versus rich input information scenarios;
\item the form of the input information on reaction directions sampled from MBE and VNC, required by Method (a) only;
\item the choice of the ``learning parameters'' $\alpha$ and $\beta$ that appear, respectively, in both methods and in Method (b) only, that characterize the size of the update step in, respectively, chemical potentials and fluxes.
\end{itemize}

\subsection{Minimal input information on chemical potentials}

To begin with, let us note that the thermodynamic constraint 
\begin{equation}\label{tc}
-\text{sign}(\nu_i)\sum_{\mu=1}^N\xi_i^\mu g^\mu\geq 0
\end{equation}
imposes that the chemical potential of metabolites that are sources (resp. sinks) of the network should be known, since the constraints (\ref{tc}) do not bound the corresponding $g^\mu$'s from above (resp. below). Clearly, then, computing the landscape of chemical potentials is feasible only if $P_0(\mathbf{g})$ carries some prior information on the $g^\mu$'s. 

Besides network `leaves', one easily sees that certain intracellular potentials should be known as well. Consider a chemical potential vector $\mathbf{g}_\star=\{g^\mu_\star\}$ that satisfies (\ref{tc}) and note that $\mathbf{g}_\star+k\boldsymbol{\lambda}$ is again a solution for $k\in\mathbb{R}$ provided $\boldsymbol{\lambda}=\{\lambda^\mu\}$ is such that $\sum_{\mu=1}^M \lambda^\mu\xi_i^\mu=0$ for each $i$. This degeneracy is related to the existence of equilibrium states and needs to be lifted. Ideally this can be achieved by fixing the chemical potentials of at least one of the metabolites with $\lambda^\mu\neq 0$ for each vector $\boldsymbol{\lambda}$ of the type described above. Note that such vectors include (but are not limited to) the conserved metabolic pools defined in \cite{famili}. A conserved pool is a group $P$ of metabolites described by a vector $\boldsymbol{\ell}=\{\ell^\mu\}$ with $\ell^\mu>0$ if $\mu\in P$, and zero otherwise, such that, for each $i$, $\sum_{\mu=1}^M\ell^\mu\xi_i^\mu =0$. From a physical viewpoint, each such pool corresponds to a conservation law for the aggregate concentration of the corresponding metabolites and the existence of one pool suffices to force $y^\mu =0$ for each $\mu \in P$ in the VNC scenario (in other words, metabolites belonging to conserved pools can not be producible) \cite{pools}. The simplest way to rule out equilibrium solutions is to clamp the chemical potential of a group $P'$ of metabolites (including network sources and sinks) such that the problem
\begin{equation}  \label{lift}
\sum_{\mu=1}^M \lambda^\mu\xi_i^\mu=\sum_{\mu \notin P'} \lambda^\mu\xi_i^\mu +\sum_{\mu \in P'} \lambda^\mu\xi_i^\mu=0~~~~~\forall i
\end{equation}
admits no solution. From a geometrical perspective, the minimal number of metabolites whose chemical potentials needs to be constrained is that for which the number of independent equations in the above system exceeds the number of variables.

\subsection{Poor versus rich input information}

In the case of the hRBC, there are four source/sink nodes, namely glucose (GLC), lactate (LAC), ammonia (NH$_3$) and carbon dioxide (CO$_2$), whereas a basis for the left kernel of the stoichiometric matrix $\boldsymbol{\Xi}$ (excluding uptakes) turns out to be composed by seven vectors: $3$ conserved  pools, namely the pairs (NAD, NADH) and (NADP, NADPH) and the larger pool formed by (HX, INO, IMP, ADE, ADO, AMP, ADP, ATP), plus $4$ other vectors whose components are related in a non-intuitive way to the balance of global quantities like the number of carbon atoms and phosphate groups. Analyzing (\ref{lift}), however, one finds that by clamping four metabolites (besides leaves) no equilibrium solution can exist.  

In the {\it poor input information} scenario, we have therefore constructed $P_0(\mathbf{g})$ by fixing the chemical potentials of GLC, LAC, NH$_3$ and CO$_2$ as well as of inosine monophosphate (IMP), adenine (ADE), NAD, and NADP to remove the degeneracy associated to the presence of equilibrium solutions (different choices for these do not alter results). For these metabolites, $P_0^\mu(g^\mu)=\delta(g^\mu-g^\mu_{\text{exp}})$, where $g^\mu_{\text{exp}}$ is the chemical potential obtained from experimental data. For the other compounds we have chosen a $P_0^\mu$ that reproduces the overall statistics of chemical potentials. In particular, each $g^\mu$ (in units of KJ/mol) is selected independently and uniformly in $[0,2000]$ with probability $p=0.2$ and in $[-5000,0]$ with probability $1-p=0.8$. 

In the {\it rich input information} scenario, $P(\mathbf{g})$ was constructed as follows. The chemical potential of metabolites whose intracellular concentrations are known experimentally were extracted from the formula
$g^\mu=g_0^\mu+RT\log c^\mu$. Here $g_0^\mu$ is the free energy of formation of the metabolite $\mu$ in the standard biochemical state. These values are accurately known for metabolites in the hRBC metabolic network (see Table \ref{metab}). The concentrations $c^\mu$ are instead taken to be uniformly and independently distributed random variables with average values and box sizes according to the experimental estimates reported in Table \ref{metab}. The concentrations of metabolites for which we were unable to find reliable experimental estimates are assumed to be uniformly and independently distributed random variables centered around $10^{-4}$ M and spanning four orders of magnitude symmetrically around the mean. Note that in this case no metabolite has a clamped chemical potential and the algorithm can modify the trial distributions of all metabolites.

In both scenarios, the chemical potential of water is treated as a boundary condition, i.e. it is kept fixed since we assume that water is in a condensed phase.

\subsection{Input information for Method (a) on directions sampled from MBE and VNC}

In \cite{palsson} and \cite{andrea}, the hRBC network is assumed to operate under the MBE and VNC scenarios, respectively, in both cases starting from prior assignments for reaction directions. The implementation of Method (a) makes use of the reaction directions obtained in these studies. In summary:
\begin{itemize}
\item according to MBE, the net flux of all reactions is in the forward direction, except PGI, R5PI and ApK, (which are found to operate bidirectionally);
\item according to VNC, the net flux of all reactions is in the forward direction, except R5PI (which is found to be operating bidirectionally), and PGI and ApK (which are found to operate in the backward direction).
\end{itemize}
For comparison the thermodynamic sampling Method (b) (which doesn't require a priori reversibility assumptions) provides solutions in which the net flux of all reactions is in the forward direction, except PGI, PGK, LDH, R5PI and ApK (which are found to operate bidirectionally). Note that reverse PGK activates a futile cycle with the Rapoport-Luebering shunt. This behavior has  been experimentally observed in acidic conditions \cite{futile}.

It should be noted however that the network used here presents an additional intracellular reaction (CA-I), an additional metabolite (HCO$_3^-$) and an additional uptake with respect to the networks studied above. Furthermore, the reconstructions employed in \cite{palsson} and \cite{andrea} have slight but important differences in the structure of uptakes.

\subsection{Setting the learning rates}

Both Methods (a) and (b) rely on learning rates (denoted as $\alpha$ for Method (a) and $\alpha$ and $\beta$ for Method (b)) that fix the size of the adjustment to be applied to chemical potentials ($\alpha$) and fluxes ($\beta$) at each iteration step.

For the implementation of Method (a) presented here, we set $\alpha$ to the (iteration-dependent) value $-2 x_{i_0}/\sum_\mu (\xi_{i_0}^{\mu})^2$. One easily sees that if we let all the chemical potentials evolve during the sampling, this value of $\alpha$ has the net effect of reverting $x_{i_0}$ at each iteration while keeping the norm $\sum_\mu (g_\mu)^2$, that defines the unit of the energy scale, constant. Note however that some of the chemical potentials must be kept fixed during the sampling (specifically those belonging to the set of compounds for which a priori information is required). In this case, the expression for $\alpha$ given above guarantees that the elementary step of the algorithm is proportional to the amount by which the constraint is violated. We found empirically that for this choice the convergence time decreases by a factor of about $10$ with respect to the case in which a constant $\alpha$ is employed. 
 
In the implementation of Method (b) we chose the value $-2y_{\mu_0}/\sum (\xi_i^{\mu_0} (\rho))^2$ for $\beta$ (for the same reasons as above), while keeping a constant learning rate for chemical potentials ($\alpha=0.001$). This was motivated by the need to try to keep a ``timescale'' separation between the dynamics of fluxes and that of chemical potentials, with the latter preferentially slower than the former. In essence, this emphasizes the role of correlations in building up the solutions while potentially limiting the exploration of states to regions in the space of chemical potentials that are not too far from the initial states. 

Finally, we remark that the solution space spanned by the completely reversible Von Neumann constraints may lack  convexity for $\rho<1$ \cite{matteofi}. This leads to a certain rejection rate of solutions (roughly 10\%), which represents a negligible additional cost in computations.

\section{Further results from Method (b)}

Fig. \ref{fig5} shows the flux-flux correlations computed from Method (b) assuming strong flux through the ATPase pump.
\begin{figure}
\begin{center}
\includegraphics*[width=0.49\textwidth,angle=0]{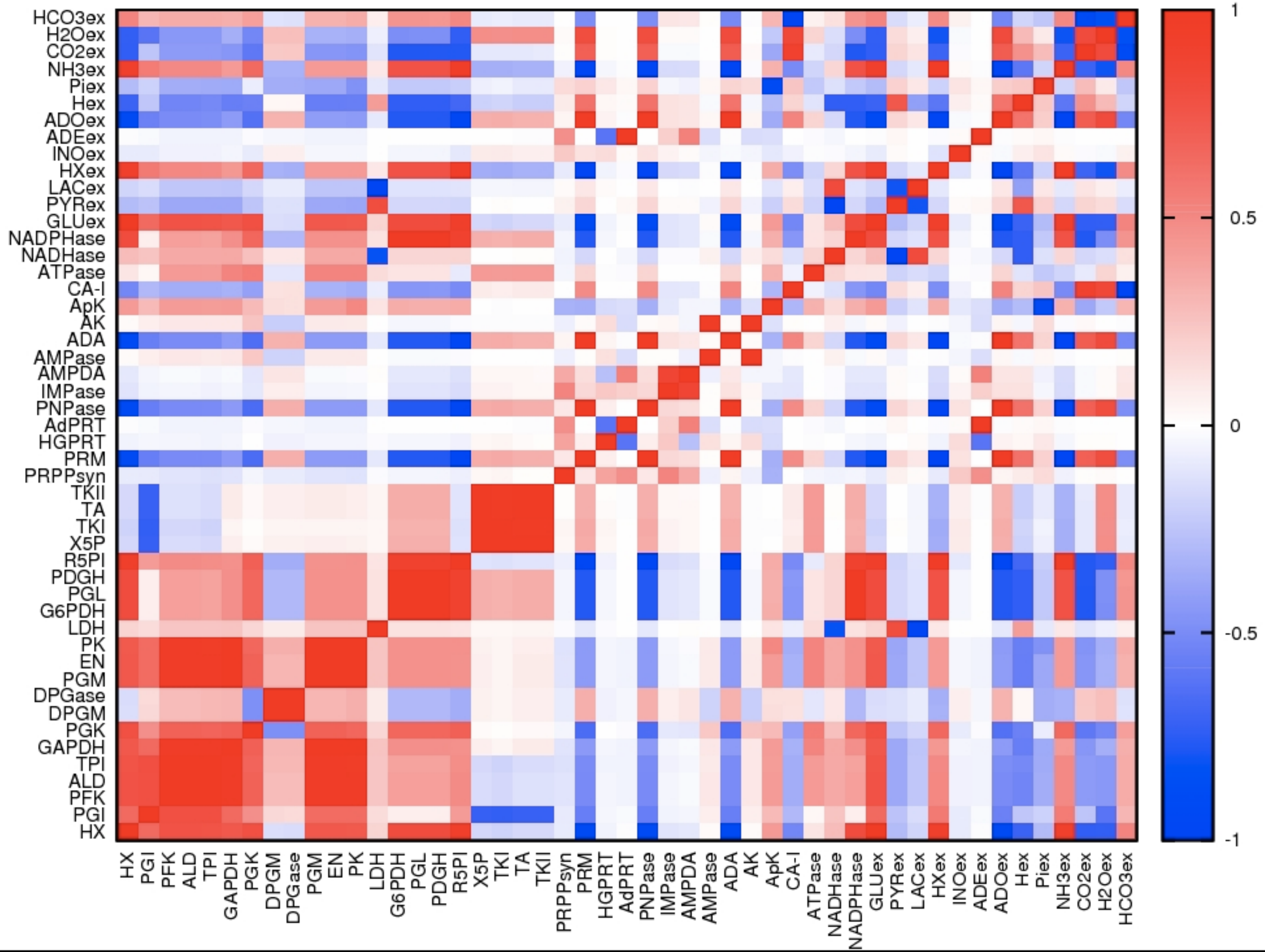}
\caption{Pearson coefficients of the fluxes obtained by Method (b) with rich input information with strong ATPase flux.}
\label{fig5}
\end{center}
\end{figure}
Compared to the case where ATPase flux is weak, one notices that glycolysis is split in two separate but correlated modules, formed respectively by the first six and the last three reactions (in agreement with \cite{palsson}). Similarly, the pentose phosphate pathway is split in two tight modules, while the nucleotides salvage pathway no longer forms a compact group. Note that the role of PGK and PGI as switches for the ATPase and NADPHase pumps is much less pronounced in this case.

\begin{figure}
\begin{center}
\includegraphics*[width=0.47\textwidth]{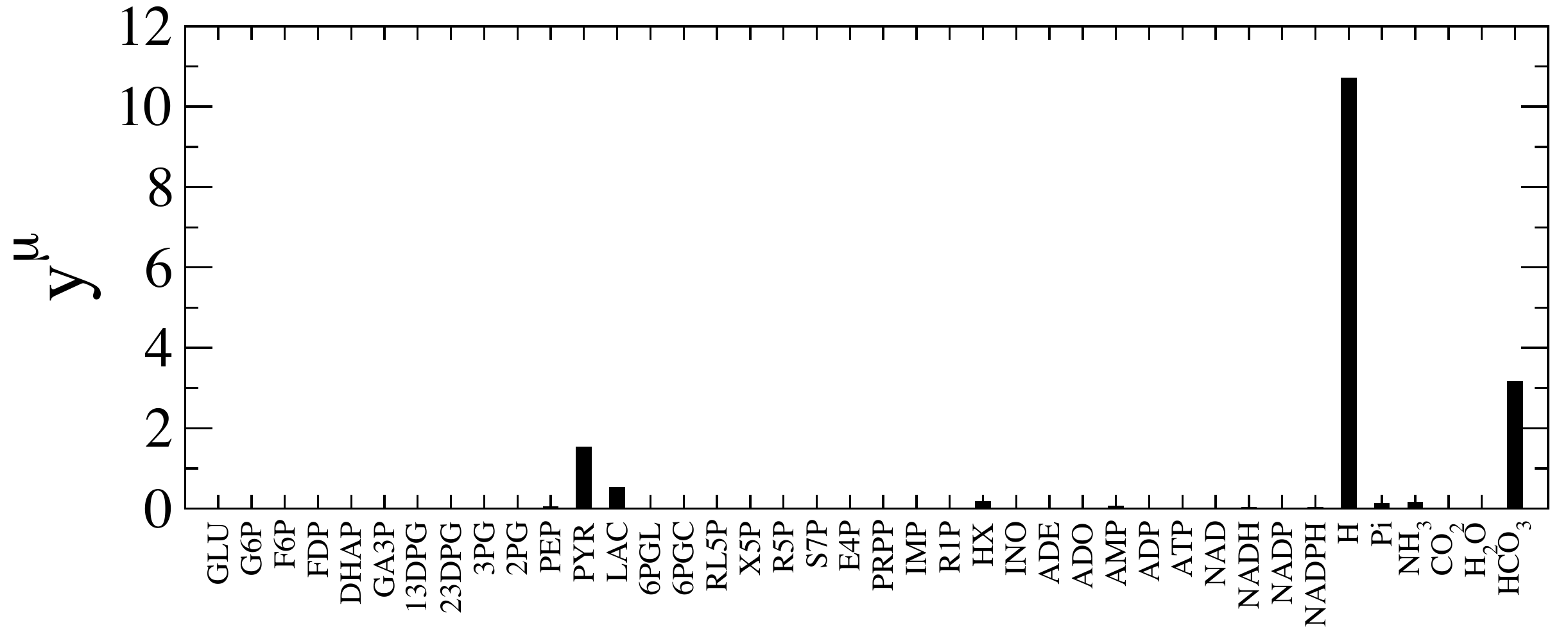}
\caption{Average production profile obtained by jointly sampling fluxes and chemical potentials with ATPase ionic pump active. Note that fluxes are measured in units of GLC uptake.}
\label{figprod}
\end{center}
\end{figure}

The average production profile computed by Method (b) is displayed in Fig. \ref{figprod}.
One sees that all intracellular metabolites are mass-balanced except PYR, LAC, H$^+$ and HCO$_3^-$. This means that VNC solutions predict a slow steady growth of their concentration. However all of these are subject to outtakes, therefore the final state of the cell is globally mass balanced (i.e. the solutions of thermodynamically-constraint VNC would coincide with those of thermodynamically-constraint MBE on the same system). 

\begin{figure}
\begin{center}
\includegraphics*[width=0.47\textwidth]{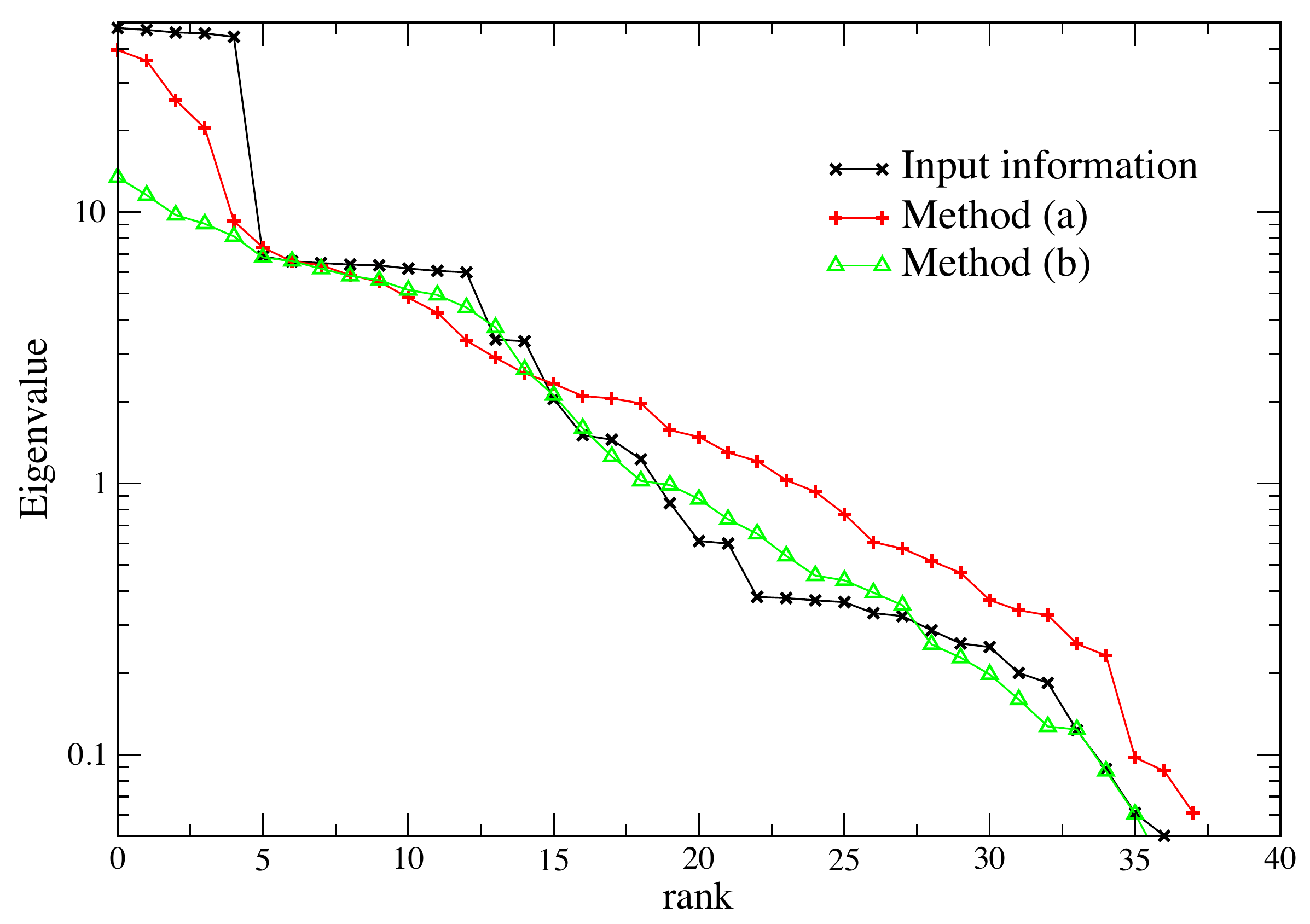}
\caption{Rank plot of the eigenvalues of the correlation matrix of chemical potentials for $P_0(\mathbf{g})$ (black), and for the distributions obtained via Methods (a) (red) and (b) (green), for rich input information. Note that the $y$-scale is logarithmic.}
\label{eigen}
\end{center}
\end{figure}

The rank plot of the eigenvalues of the chemical potential correlation matrix $\mathbf{K}$ computed from Methods (a) and (b) is reported in Fig. \ref{eigen}.
One clearly sees that Method (b) outperforms Method (a) in gaining information over the trial distribution $P_0(\mathbf{g})$. Note in particular that Method (a) typically loses information on metabolites with small variability range in the trial distribution (corresponding to the smallest eigenvalues of $\mathbf{K}$).

Finally, Table \ref{table} displays a summary of the chemical energy balance of the hRBC metabolism. 
\begin{table}
\begin{center}
\begin{tabular}{| c |c | c | c | c | }
\hline
State & $T\dot{S}$(fW) & $\dot{W}$(fW) & $\dot{U}$(fW) & $\eta$ \\
\hline
average eff., weak ATPase flux & 226 & 51 & 278 & 0.18 \\
average eff., strong ATPase flux & 235 & 67 & 302 & 0.22 \\
maximal eff., weak ATPase flux & 256 & 92 & 348 & 0.264 \\
maximal eff., strong ATPase flux & 268 & 112 & 381 & 0.295 \\
\hline
\end{tabular}
\caption{Summary of the energy balance of the human red blood cell metabolic network}
\label{table}
\end{center}
\end{table}

\section{Approximate energy balance analysis.}

One of the results emerging from the sampling of fluxes and chemical potentials is that the overall flux through the nucleotide rescue pathway is about one order of magnitude lower than that going through glycolysis and/or pentose phosphate pathway. This is in agreement with previous results based on MBE and VNC without explicit thermodynamic constraints \cite{andrea} and the known experimental values (see Table \ref{tableflux}).
\begin{table}
\begin{center}
\begin{tabular}{| c |c | }  
\hline 
Reaction & Flux [M/s] \\ 
\hline 
HK  & $4.3 \pm 0.7 \cdot 10^{-6}$ \cite{tornbum} \\
GSSGR & $5 \pm 1 \cdot 10^{-5}$ \cite{tornbum} \\
RLS & $1.4 \cdot 10^{-7}$ \cite{rapoport1} \\
AMPcat & $3 \cdot 10^{-9}$ \cite{rapoport1} \\
\hline
\end{tabular}
\caption{Measured fluxes in the red blood cell. GSSGR stands for glutathione reductase; RLS for Rapoport-Luebering shunt; AMPcat is AMP catalysis.}
\label{tableflux}
\end{center}
\end{table}

It is interesting to notice that in order to maintain the $ase$ pumps the activation of glycolysis and pentose phosphate pathway is sufficient. To see this, we can work out the stationary state of the network in absence of the nucleotide salvage pathway, i.e. for a reduced network consisting of 13 (glycolysis) + 8 (PPP) + 3 (pumps) = 24 intracellular reactions and 5 uptakes (GLC, PYR, LAC H$_2$O and HCO$_3$). The mass balance equations reveal that only four of the above reactions are linearly independent: we choose $u_{\text{GLC}}$, $u_{\text{LAC}}$ (glucose and lactate uptake), $\nu_{\text{G6PDH}}$ and $\nu_{\text{RLS}}$  (RLS stands for Rapoport-Luebering shunt). All fluxes can be written in terms of these. 
In particular, the quantities of interest in the energy balance take the form
\begin{multline}
\dot{U} = u_{\text{GLC}} (g^{\text{GLC}}_{(ext)} -2 g^{\text{PYR}}_{(ext)}) + u_{\text{LAC}} (g^{\text{LAC}}_{(ext)} - g^{\text{PYR}}_{(ext)}) +\\+\nu_{\text{G6PDH}}(1/3 g^{\text{PYR}}_{(ext)} + 2g^{\text{H$_2$O}}_{(ext)} - g^{\text{HCO$_3$}}_{(ext)}) 
\end{multline}
for the energy flow and
\begin{gather}
\nu_{\text{ATPase}} = 2 u_{\text{GLC}} -\nu_{\text{G6PDH}}/3 - \nu_{\text{RLS}} \\
\nu_{\text{NADHase}} = 2 u_{\text{GLC}} - \nu_{\text{G6PDH}} +u_{\text{LAC}} \\
\nu_{\text{NAPDHase}} = \nu_{\text{GSSGR}} = 2 \nu_{\text{G6PDH}}
\end{gather}
for the fluxes through the pumps. Now under flux balance $u_{\text{GLC}}=\nu_{\text{HK}}$. Hence from the values in Table \ref{tableflux} we can estimate $\nu_{\text{ATPase}} \simeq 3 \cdot 10^{-7}$ M/s, i.e. one order of magnitude smaller than the glucose uptake
as in \cite{andrea} and in agreement with our thermodynamic sampling without constraints on the ATP ionic pump.

\end{document}